\documentclass[twocolumn]{aastex63}
\usepackage{graphicx,float,subfigure} %including figures/graphics
\usepackage{booktabs}
\usepackage{float}
\usepackage[T1]{fontenc}
\usepackage[utf8]{inputenc}
\usepackage{ae}
\usepackage{aecompl}
\usepackage{amsmath, amssymb, amsfonts, braket, gensymb} %math symbols/writing
\usepackage{newtxtext,newtxmath} %Times  font
\usepackage{enumitem}
\usepackage{longtable}
\def\url#1{\expandafter\string\csname #1\endcsname}

\shorttitle{Red Clump Color-Metallicity Relation}
\shortauthors{Nataf et al.}

\begin{document}
	
\title{On the Color-Metallicity Relation of the Red Clump and the Reddening Toward the Magellanic Clouds}
\correspondingauthor{David M. Nataf}
\email{dnataf1@jhu.edu, david.nataf@gmail.com}

\author{David M. Nataf}
\affil{Center for Astrophysical Sciences and Department of Physics and Astronomy,
The Johns Hopkins University, 
Baltimore, MD 21218}
\author{Santi Cassisi}
\affiliation{INAF-Osservatorio Astronomico d'Abruzzo, via M. Maggini, sn. I-64100 Teramo, Italy}
\affiliation{INFN-Sezione di Pisa, Largo Pontecorvo 3, I-56127 Pisa, Italy}
\author{Luca Casagrande}
\affiliation{Research School of Astronomy and Astrophysics, Mount Stromlo Observatory, The Australian National University, ACT 2611, Australia}
\author{Wenlong Yuan}
\affiliation{Center for Astrophysical Sciences and Department of Physics and Astronomy,
The Johns Hopkins University,
Baltimore, MD 21218}
\author{Adam G. Riess}
\affiliation{Space Telescope Science Institute, 3700 San Martin Drive, Baltimore, MD 21218, USA}
\affiliation{Center for Astrophysical Sciences and Department of Physics and Astronomy,
The Johns Hopkins University,
Baltimore, MD 21218}
\begin{abstract}
	The zero point of the reddening toward the Large Magellanic Cloud (LMC) has been the subject of some dispute.  Its uncertainty propagates as a systematic error for methods which measure the extragalactic distance scale through knowledge of the absolute extinction of LMC stars. In an effort to resolve this issue, we used three different methods to calibrate the most widely-used metric to predict LMC extinction, the intrinsic color of the red clump, $(V-I)_{RC,0}$, for the inner $\sim$3 degrees of that galaxy. The first approach was to empirically calibrate the color zeropoints of the BaSTI isochrones over a wide metallicity range of ${\Delta}\rm{[Fe/H]} \approx 1.10$ using measurements of red clump stars in 47 Tuc, the Solar Neighborhood, and NGC 6791.
	From these efforts we also measure these properties of the Solar Neighborhood red clump, ($V-I$, $G_{BP}-K_{s}$, $G-K_{s}$, $G_{RP}-K_{s}$, $J-K_{s}$, $H-K_{s}$, $M_{I}$, $M_{Ks}$)$_{RC,0} =$ (1.02, 2.75, 2.18, 1.52, 0.64, 0.15, $-$0.23, $-$1.63). The second and third methods were to compare the observed colors of the red clump to those of Cepheids and RR Lyrae in the LMC %which are standardizable crayons based on their periods and metallicities. 
		With these three methods, we estimated the intrinsic color of the red clump of the LMC to be $(V-I)_{RC,0,\rm{LMC}} = \{ \approx 0.93,0.91 \pm 0.02,0.89 \pm 0.02\}$ respectively, and similarly using the first and third method we estimated $ (V-I)_{RC,0,\rm{SMC}} = \{\approx 0.85,0.84 \pm 0.02 \}$ respectively for the Small Magellanic Cloud. We estimate the luminosities to be $M_{I,RC,\rm{LMC}}=-0.26$ and $M_{I,RC,\rm{SMC}}=-0.37$. We show that this has important implications for recent calibrations of the tip of the red giant branch in the Magellanic Clouds used to measure $H_0$.
\newline\vspace{2.0em}
\end{abstract}
 %These three stellar populations span a metallicity range of ${\Delta}\rm{[Fe/H]} \approx 1.10$, and thus have great leverage as a calibration set.
	%$(G_{BP}-K_{s})=2.75$, $(G-K_{s})=2.18$, $(G_{RP}-K_{s})=1.52$, and $(H-K_{s})=0.15$.
	%$((V-I)_{RC,0},(J-K_{s})_{RC,0},M_{I,RC},M_{Ks,RC}) = (1.02,0.64,-0.23,-1.63)$.
%The Cepheid data is newly available from {\it HST} WFC3 and has greater photometric precision than past ground-based data used for this purpose.
	
\section{Introduction} \label{sec:Introduction}

The red clump, which is the horizontal branch of a metal-rich ([Fe/H] $\gtrapprox -1.0$) stellar population, is a long-lived and thus well populated evolutionary stage and predicted by stellar theory to have precise color-metallicity and luminosity-metallicity relations \citep{2001MNRAS.323..109G,2002MNRAS.337..332S}. The red clump has thus been a widely used ``standard crayon" and standard candle to characterize stellar populations over a wide range of ages, for example the mapping of the chemical gradients of the Milky Way Disk \citep{2014ApJ...796...38N}; mapping the reddening toward the Milky Way's nuclear disk \citep{2019NatAs.tmp....4N}; the reddening variations \citep{2016MNRAS.456.2692N}, bar morphology \citep{1997ApJ...477..163S}, peanut/X-shape profile  \citep{2010ApJ...721L..28N,2010ApJ...724.1491M} and the long bar component \citep{2015MNRAS.450.4050W} of the Milky Way Bulge;  the reddening toward the Magellanic Clouds \citep{2011AJ....141..158H,2020ApJ...889..179G}; the reddening of the Tip of the Red Giant Branch (TRGB) for distance scale work \citep{2017ApJ...835...28J,2017ApJ...836...74J, 2019ApJ...886...61Y}, the reddening of and distance to the Sagittarius stream \citep{2010ApJ...721..329C,2012AJ....144...18C} and the Andromeda galaxy \citep{1998ApJ...503L.131S,2018AJ....156..230C}. 

That is an impressive breadth of discovery and diagnostic power, and one which we expect may grow in the coming decade as several new observatories come online. Each of the \textit{James Webb Space Telescope (JWST)}, The \textit{Wide-Field Infrared Survey Telescope (WFIRST)}, the \textit{Giant Magellan Telescope (GMT)}, the \textit{Thirty Meter Telescope (TMT)}, and the \textit{European Extremely Large Telescope (E-ELT)} \citep{2007Msngr.127...11G} will be maximally sensitive for wavelengths with $\lambda \gtrapprox 1 {\mu}$m, a regime where red clump stars are intrinsically bright due to their cool ($T_{\rm{eff}} \approx 4,500\,$K) effective temperatures. For example, $M_{\rm{RC,Ks}} \approx -1.62$ for the stellar population of the solar neighbourhood \citep{2017MNRAS.471..722H,2019arXiv191000398C,2017ApJ...840...77C}. 

We have thus been motivated to explore a discrepancy in the assumed intrinsic red clump optical color $(V-I)_{RC,0}$ in the recent literature. The assumption directly affects the reddening toward the Large Magellanic Cloud (LMC), for which the widely-used reddening map of \citet{2011AJ....141..158H} had assumed $(V-I)_{RC,0,\rm{LMC}}=0.92$. This assumption is corroborated by other methods. For example, the reddening map of \citet{2016ApJ...832..176I}, which is derived from measurements of classical Cepheids, estimated a reddening toward the LMC that was 0.02 mag greater in the mean, an offset smaller than the reported errors. However, the recent investigation of \citet{2020ApJ...889..179G} derived a much bluer value, of $(V-I)_{RC,0,\rm{LMC}} = 0.838 \pm 0.034$ for the LMC. Stellar population gradients might also be an issue. Indeed, the investigation of \citet{2020arXiv200602448S}, which was submitted at around the same time as our own, derived $(V-I)_{RC,0,\rm{LMC}} = 0.8915 - d \times 0.0025$, where $d$ is the projected separation (in degrees) of a field with respect to the center of the LMC.  In this work we focus on the properties of the red clump in the inner few degrees of the LMC and SMC. 

That offset, perhaps as large as 0.08 mag in $(V-I)_{RC,0,\rm{LMC}}$, is substantially larger than the reported errors of most, and indeed perhaps all, recent investigations that assume an intrinsic color for the red clump. It may be due to factors such as undiagnosed population effects in the LMC stellar population, or the red clump color-metallicity relation having a very different slope than commonly assumed. Either finding would necessitate a revision of much of the literature on stellar astrophysics, interstellar reddening, and the distance scale. It also might have a simpler explanation if the sight lines of young, blue supergiants used by \citet{2020ApJ...889..179G} to deredden red clump stars are dustier than those of the older, red giants which we consider in Section \ref{sec:Discussion}.  

The purpose of this work is to investigate the issue of the red clump's color-metallicity relation, and to derive robust zeropoints for the red clump for use in the LMC and the Small Magellanic Cloud (SMC). We first review the theoretical and empirical calibration of the red clump in Section \ref{sec:Calibration}. We verify the consistency of these estimates for the LMC and SMC in Section \ref{sec:Consistency}. In Section \ref{sec:Analysis}, we evaluate the consistency of the predicted value of $(V-I)_{RC,0}$ for the LMC and SMC with calibrations from exquisitely-measured Cepheids and RRab stars. We discuss our findings and present our conclusions in Sections \ref{sec:Discussion} and \ref{sec:Conclusion}.

%The red clump has historically been a widely used standard crayon for globular clusters, the bulge, the magellanic clouds, andromeda, and will soon be used to greater distances with JWST, GMT, TMT, WFIRST, etc.  Refer to classic papers by Salaris and Girardi. 

%Recent discrepancy in the literature on LMC+SMC, Haschke et al (2011) derive a reddening map assuming (V-I)0=0.92. However, Gorski et al. (2020) derive (V-I)0=0.84 by assuming that red clump stars are extinguished by the same amount as nearby OB stars. 

%We show that this assumption is erroneous, and thus that the red clump color-metallicity relation is precise over a span of ~1 dex in metallicity, that LMC and SMC reddening values can be precisely assumed, and that OB stars should thus be assumed to be more extinguished than nearby red clump stars. 
\section{Calibration of the Red Clump} \label{sec:Calibration}

In this Section we will employ a hybrid empirical and theoretical approach to predict the color and magnitude of the red clumps of the Large and Small Magellanic Clouds. We will follow the approach of \citet{2001MNRAS.323..109G} and \citet{2002MNRAS.337..332S}, in which the trends of red clump color and magnitude with age and metallicity are derived from stellar models, but the zero points are shifted to agree with the best available empirical data. 

The red clumps of the globular cluster 47 Tuc, of the solar neighbourhood, and of the open cluster NGC 6791 will be used to calibrate the zero points. For each of those three stellar populations, and both Magellanic Clouds, we will state both our assumed population parameters ([Fe/H], [$\alpha$/Fe], etc), as well as other robust estimates from the literature, as the uncertainty in population parameters contributes to the uncertainty in the overall calibration. 

For the chemical abundances ([Fe/H], [$\alpha$/Fe]) we assume the parameters derived by the ASPCAP pipeline \citep{2016AJ....151..144G} from spectra taken for APOGEE \citep{2017AJ....154...94M}, a high-resolution ($R \sim 22,500$, high signal-to-noise (SNR $\sim$ 100) Galactic archaeology spectroscopic survey that is part of the Sloan Digital Sky Survey \citep{2017AJ....154...28B}. This survey is chosen as it has large numbers of consistent measurements of the stellar populations studied in this work. The [Fe/H] abundances are consistent with prior literature values, with a mean and standard deviation in the offset with respect to 525 ``standard stars" of ${\Delta}\rm{[Fe/H]} = 0.04 \pm 0.10$ dex, and is largely a uniform shift, without a metallicity trend (Table 4 and Figure 3 of \citealt{2018AJ....156..126J}, respectively).  We also verify that this is the case for the particular populations studied in this work. We use these abundances not just for the LMC and SMC, but for all stellar populations discussed in this work, to ensure a homogeneous metallicity scale.

\subsection{Population Parameters for the Large and Small Magellanic Clouds} \label{subsec:MCpop}

We assume the star-formation histories measured by \citet{2013MNRAS.431..364W}, which can be read off their Figures 4 and 5 and were also kindly sent to us by email. These star-formation histories were derived by comparing deep color-magnitude diagrams observed with \textit{HST}, with synthetic color-magnitude diagrams created by MATCH \citep{2002MNRAS.332...91D}. The resulting mean ages for the LMC and SMC stellar populations are 6 and 5 Gyr respectively. 

Seven of the eight LMC fields, and three of the four SMC fields, studied by \citet{2013MNRAS.431..364W} are toward sightlines closer to the centers of those galaxies than the majority of the tracers we use in this work. Based on their Figure 4, we may thus be underestimating the age by $\sim$1 Gyr. As we will show in Section \ref{subsec:Theory}, this would have a negligible effect on the derived red clump color, and a 0.03 mag effect on the predicted red clump brightness. 

Regardless of how weigh over the different star-formation histories, these age estimates are not consistent with several prior literature results. A significant contributor to this difference is that the mean age of the stellar population need not be the same as the mean age of stars in the red clump region of the color-magnitude diagram, a correction that we explain and account for in Section \ref{subsec:Composite}. For example,  \citet{2018MNRAS.478.5017R} derived $\tau_{\rm{SMC}} \approx 3$ Gyr for stars in and near the red clump (Region D panel of their Figure 13), and \citet{2018ApJ...866...90C} derived $\tau_{\rm{LMC}} \approx 1.6$ Gyr by assuming [Fe/H]$=-0.50$ and the age-metallicity relation of \citet{2008AJ....135..836C}. \citet{2017ApJ...847..102Y} did derived a $\sim$2 Gyr mean age for the red clump of the SMC using a more comprehensive color-magnitude diagram.

On the other hand, some prior literature estimates were more consistent with ours, or even derived an older age. \citet{2013AJ....145...17P} estimated the ages and metallicities of LMC red giants, from which they derive $\tau=5$ Gyr at [Fe/H]$-0.70$, their mean LMC metallicity for stellar populations at projected separations similar to those studied in this work. \citet{2020ApJ...895...88N} fit a chemical evolution model to LMC abundance data, from which they derived a mean age of $\tau \approx$ 10 Gyr for the LMC.

%as they estimate the ages of stars are ultimately inferred from their positions on the red giant branch, and thus have uncertainties that are more strongly degenerate with the assumed values of distance, extinction, and metallicity than the method of \citet{2013MNRAS.431..364W}, which included each of main-sequence, subgiant branch, and post-main-sequence stars. See also \citet{2017ApJ...847..102Y}.

We use the abundances of faint red giants and bright red giants, as defined by \citet{2020ApJ...895...88N} for the APOGEE LMC fields of LMC\{4,5,6,8,9,13,14\} and SMC fields of SMC\{3,4,5\}. We exclude stars with $\log{g} \geq 2$ as likely being foreground contaminants, and stars with [Fe/H]$\leq -1.1$ as likely not contributing to the red clump. This yields a sample of 882 red giants in the LMC, and 310 red giants in the SMC. The resulting mean values of $( \langle \rm{[Fe/H]} \rangle, \langle [\alpha/\rm{Fe}] \rangle)$ are $(-0.64,+0.03)$ and $(-0.88,-0.02)$ for the LMC and SMC respectively, with a mean angular separation from the centers of their host galaxies of 2.8 and and 1.6 degrees respectively.  A cut at [Fe/H]$\leq -1.5$ would shift the effective mean metallicities by 0.04 and 0.08 dex for the LMC and SMC respectively. The mean metallicity derived by the investigation of \citet{2013MNRAS.431..364W} is nearly identical, for example  $( \langle \rm{[Fe/H]}_{\rm{LMC}} \rangle  =-0.63$.

These are consistent with other literature values. From the work of  \citet{2013A&A...560A..44V}, we find that the mean abundance of 148 LMC bar and disk stars satisfying [M/H] $\geq -1.1$ is $\langle\rm{[M/H]}\rangle = -0.64$. Similarly, from the work of \citet{2016AJ....152...58P} we find an abundance of $\langle\rm{[Fe/H]}\rangle = -0.81$ for the SMC, and \citet{2014AN....335...79M} reported $\langle [\alpha/\rm{Fe}] \rangle \approx +0.10$.

%\citet{2014AN....335...79M} measured $(\langle\rm{[Fe/H]}\rangle, \langle [\alpha/\rm{Fe}] \rangle) \approx (-0.9, +0.1)$ for 200 red giants in the SMC, and \citet{2016AJ....152...58P} measured a median metallicity of [Fe/H]$=-$0.97 from a sample of 750 calcium triplet measurements. 

The full list of adopted population parameters, for all stellar populations discussed in this work, can be found in Table \ref{table:pop_params}.

\subsection{Theoretical Calibration of the Red Clump} \label{subsec:Theory}

For the theoretical predictions of the color of the red clump, we use the isochrones from the BaSTI database\footnote{\url{http://basti.oa-abruzzo.inaf.it/index.html}}, for both the scaled-solar and $\alpha$-enhanced ([$\alpha$/Fe]$=+0.40$) cases \citep{2004ApJ...612..168P,2006ApJ...642..797P,2007AJ....133..468C}. We use the models with the suggested opacities \citep{2005ApJ...623..585F} and with red giant mass loss parameter of $\eta = 0.20$ \citep{1975MSRSL...8..369R}. This value has been validated by \citet{2012MNRAS.419.2077M} in their study of NGC 6791 and 6819, two open clusters for which there are asteroseismic data. They reported that the data were consistent with a mass loss parameter of $0.10 \lesssim \eta \lesssim 0.30$. We use isochrones with the metallicities [M/H] $= \{-0.66,-0.35,-0.25,+0.06,+0.26,+0.40\}$ and ages $\tau$/Gyr $=\{2.5,\,5.0,\,10.0,\,12.0\}$, we also include isochrones with ([Fe/H],[$\alpha$/Fe])$=(-0.96,0.00)$. 

The predicted mean color and mean magnitude of the red clump are derived by computing the weighted mean of all equivalent evolutionary points that are no brighter than 0.07 mag brighter than the red clump, and that correspond to evolutionary stages subsequent to the helium flash.  Adjusting the 0.07 mag criterion by 0.01 or 0.02 mag has a negligible impact ($ \approx 0.001$ mag) on the derived predicted color and magnitude of the red clump. The full list of predicted magnitudes and colors can be found in Table \ref{table:BaSTI}, and are plotted in Figure \ref{figure:BaSTIpred}. 
{
\begin{figure}
    \includegraphics[width=1.0\linewidth]{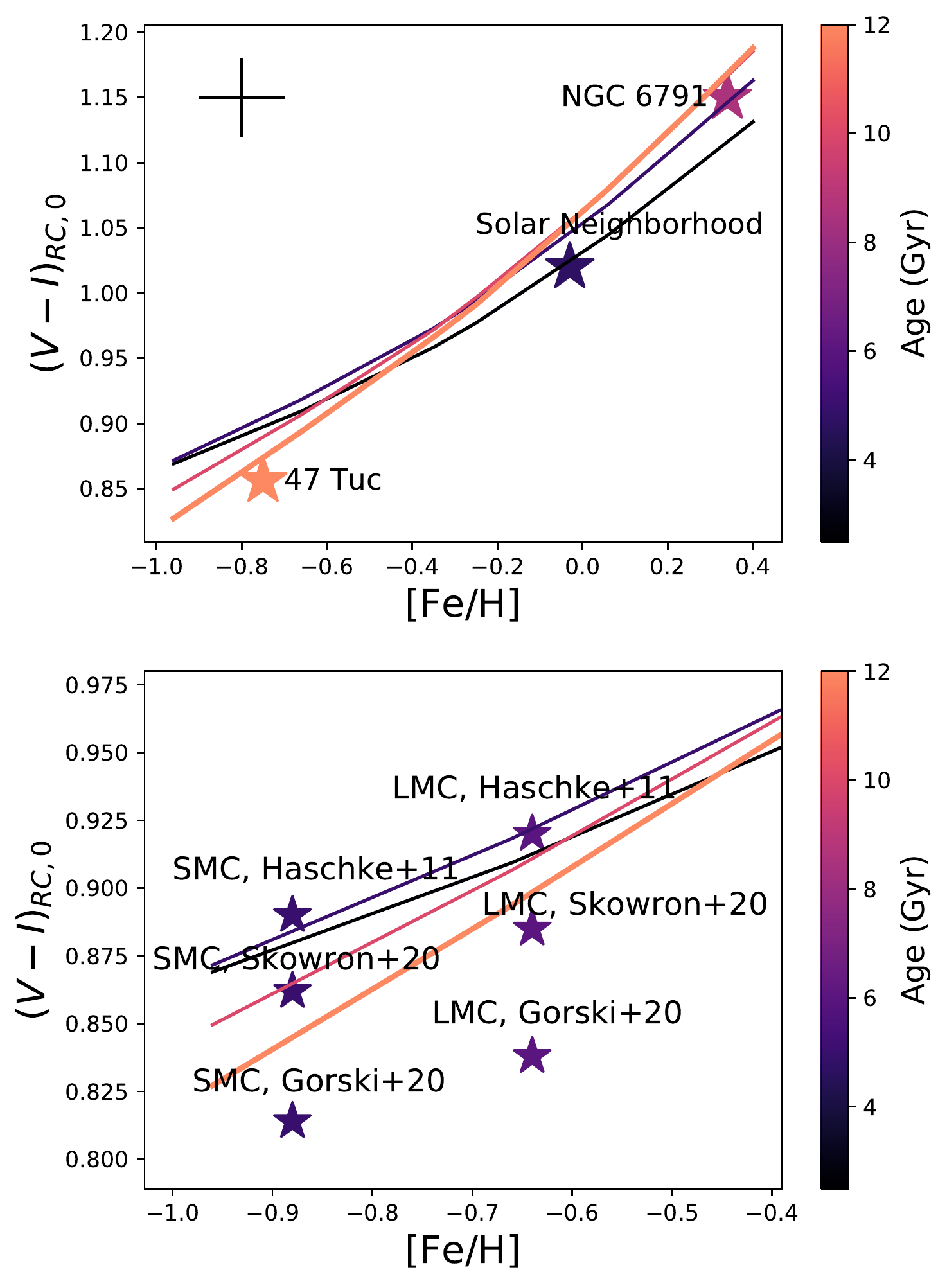}
    \caption{The BaSTI-predicted values of $(V-I)_{RC,0}$ as a function of [Fe/H] for [$\alpha$/Fe]$=0$ and $\tau/\rm{Gyr}=(2.5,5.0,10.0,12.0)$, shown as the curves that are color-coded by age. 
    We show the $(V-I)_{RC,0}$ values for three calibrating stellar populations as asterisks color-coded for ages, with a characteristic error bar shown in the top-left of the plot. The adopted values of [Fe/H], age, etc are justified in Section  \ref{subsec:Empirical}. The $(V-I)_{RC,0}$ of 47 Tuc is shifted by 0.025 mag due to its elevated $\alpha$-element and initial helium abundance. The measured values of $(V-I)_{RC,0}$ are on average $0.03 \pm 0.01$ mag lower (bluer) than the BaSTI-predicted values.  A zoom-in of the lower-metallicity relationship is shown in the bottom panel, where we also show, as asterisks color-coded for age, the $(V-I)_{RC,0}=0.89,0.92$ values assumed by \citet{2011AJ....141..158H}, the $(V-I)_{RC,0} \approx 0.885,0.862$ values assumed by  \citet{2020arXiv200602448S} at the mean position of the red giant stars with spectroscopic metallicities used as calibrators here, and the  $(V-I)_{RC,0}=0.814 \pm 0.034,0.838 \pm 0.034$ values assumed by \citet{2020ApJ...889..179G}, with [Fe/H] and age assumptions justified in Section \ref{subsec:MCpop}. 
    %The BaSTI-predicted values of $(V-I)_{RC,0}$ (TOP) and $M_{I,RC}$ (BOTTOM) as a function of [Fe/H] for [$\alpha$/Fe]$=0$, shown as points that are color-coded by age. We show, as color-coded X's, the values for three stellar populations used as calibrators in this work, and we also show, as upwards triangles, the $(V-I)_{RC,0}$ values assumed by \citet{2011AJ....141..158H} and \citet{2020ApJ...889..179G}. The adopted values of [Fe/H], age, etc are justified in Section  \ref{subsec:Empirical}. Metal-rich red clumps are predicted typically redder and fainter, and younger red clumps are predicted to be brighter. The measured values are on average $\sim$0.02 mag bluer, and $\sim$0.10 mag brighter, than the predicted values.
    \vspace{1em}}
    \label{figure:BaSTIpred}
\end{figure}

We find that the dependence of magnitude and color on population parameters cannot be adequately fit for with a simple function over the full parameter space. For example, the dependence of $(V-I)_{RC,0}$ on age is itself a function of metallicity, see Figure \ref{figure:BaSTIpred}. In order to estimate the prediction for the LMC directly from the isochrones, without first deriving empirically-calibrated offsets, we compute a linear fit over a limited parameter range, with [Fe/H] $\leq -0.25$, $2.5 \leq \tau/\rm{Gyr} \leq 10$:

%LMC, VMI: [0.92075371 0.19927221 0.12171493 0.01026804]
\begin{equation}
\begin{split}
        (V-I)_{RC,0,\rm{LMC}} \approx 0.92 + 0.20\bigl(\rm{[Fe/H]}_{\rm{LMC}}+0.64\bigl)  + \\  0.12\bigl(\rm{[\alpha/\rm{Fe}]}_{\rm{LMC}} -0.03\bigl) + 0.01 \log \bigl( \tau_{\rm{LMC}}/\rm{6\,Gyr} \bigl)
    \label{EQ:VMI_Basti}
     \end{split}
\end{equation}
%LMC, MI: [-0.42116134  0.18906295  0.05605885  0.36198894]
\begin{equation}
\begin{split}
    M_{I,RC,\rm{LMC}} \approx -0.42 + 0.19\bigl(\rm{[Fe/H]}_{\rm{LMC}}+0.64\bigl) + \\ 0.06\bigl(\rm{[\alpha/\rm{Fe}]}_{\rm{LMC}}-0.03\bigl) + 0.36 \log \bigl( \tau_{\rm{LMC}}/\rm{6\,Gyr} \bigl),
    \label{EQ:MI_Basti}
     \end{split}
\end{equation}
where the x-intercepts are chosen to correspond to the population parameters of the LMC. Similarly for the SMC, we compute a linear fit over a limited parameter range, with [Fe/H] $\leq -0.60$, $\tau/\rm{Gyr} \leq 10$:
%SMC, VMI: [ 0.87371493  0.17772253  0.09482243 -0.00474685]
\begin{equation}
\begin{split}
        (V-I)_{RC,0,\rm{SMC}} \approx 0.87 + 0.18\bigl(\rm{[Fe/H]}_{\rm{SMC}}+0.88\bigl)  + \\  0.09\bigl(\rm{[\alpha/\rm{Fe}]}_{\rm{SMC}} +0.02\bigl) + 0.00 \log \bigl( \tau_{\rm{SMC}}/\rm{5\,Gyr} \bigl)
    \label{EQ:VMI_Basti2}
     \end{split}
\end{equation}
%SMC, MI: [-0.50023003  0.14594949  0.07180636  0.38643235]
\begin{equation}
\begin{split}
    M_{I,RC,\rm{SMC}} \approx -0.50 + 0.15\bigl(\rm{[Fe/H]}_{\rm{SMC}}+0.88\bigl) + \\ 0.07\bigl(\rm{[\alpha/\rm{Fe}]}_{\rm{SMC}}+0.02\bigl) + 0.39 \log \bigl( \tau_{\rm{SMC}}/\rm{5\,Gyr} \bigl),
    \label{EQ:MI_Basti2}
     \end{split}
\end{equation}
 The scatter between the model-predicted values and the fit-predicted values is $\sim$0.01 mag. 
 
 The predicted theoretical dependence of $(V-I)_{RC,0}$ on age and metallicity is plotted in Figure \ref{figure:BaSTIpred}, where we also show the empirical measurements for our three calibrating populations, and the previously assumed parameters for the LMC and SMC.

%The predictions for the $\alpha$-enhanced isochrones can be compared to those of the scaled-solar isochrones by using a bulk metallicity as defined \citep{1993ApJ...414..580S}, $\rm{[M/H]}=\rm{[Fe/H]}+\log_{10} \{ {0.638*10^{[\alpha/\rm{Fe}]}+0.362} \}$. That does not quite work -- the resulting fit predicts values of $(V-I)_{RC,0}$ and $M_{RC,I}$ that are reduced by 0.04 and 0.02 mag respectively. In other words, the effect of ${\Delta}$[$\alpha$/Fe] is in this case \textit{smaller} than expected from the formula of \citet{1993ApJ...414..580S}. 

We did verify how increasing the color range to the red to include the first ascent red giant branch would shift the derived colors. That is in principal more appropriate for composite populations such as the LMC, the SMC, and the Solar Neighbourhood, for which the red clump is inevitably mixed with the first ascent red giant branch. We found that the effect is small -- the predicted $(V-I)_{RC,0}$ is increased by 0.005 mag. That is because of the relative number counts of the two phases. \citet{2014MNRAS.442.2075N} computed, using the same isochrones as used here, that the equivalent width of the red clump above the underlying red giant luminosity function is approximately 2.0 magnitudes.

\begin{table}
\centering 
\renewcommand{\arraystretch}{0.85}
\begin{tabular}{| lll | cc |} 
\hline
[Fe/H] & [$\alpha$/Fe] & $\tau$/Gyr & M$_{I,RC}$ & $(V-I)_{RC,0}$ \\
\hline
$-$1.01 & 0.40 & 2.50 & $-$0.61 & 0.89 \\ 
$-$1.01 & 0.40 & 5.00 & $-$0.51 & 0.89 \\ 
$-$1.01 & 0.40 & 10.00 & $-$0.37 & 0.88 \\ 
$-$1.01 & 0.40 & 12.00 & $-$0.31 & 0.86 \\ 
$-$0.96 & 0.00 & 2.50 & $-$0.65 & 0.87 \\ 
$-$0.96 & 0.00 & 5.00 & $-$0.53 & 0.87 \\ 
$-$0.96 & 0.00 & 10.00 & $-$0.36 & 0.85 \\ 
$-$0.96 & 0.00 & 12.00 & $-$0.29 & 0.83 \\ 
$-$0.70 & 0.40 & 2.50 & $-$0.55 & 0.94 \\ 
$-$0.70 & 0.40 & 5.00 & $-$0.46 & 0.95 \\ 
$-$0.70 & 0.40 & 10.00 & $-$0.33 & 0.94 \\ 
$-$0.70 & 0.40 & 12.00 & $-$0.29 & 0.94 \\ 
$-$0.66 & 0.00 & 2.50 & $-$0.59 & 0.91 \\ 
$-$0.66 & 0.00 & 5.00 & $-$0.48 & 0.92 \\ 
$-$0.66 & 0.00 & 10.00 & $-$0.34 & 0.91 \\ 
$-$0.66 & 0.00 & 12.00 & $-$0.29 & 0.89 \\ 
$-$0.60 & 0.40 & 2.50 & $-$0.52 & 0.96 \\ 
$-$0.60 & 0.40 & 5.00 & $-$0.43 & 0.97 \\ 
$-$0.60 & 0.40 & 10.00 & $-$0.32 & 0.97 \\ 
$-$0.60 & 0.40 & 12.00 & $-$0.28 & 0.96 \\ 
$-$0.35 & 0.00 & 2.50 & $-$0.50 & 0.96 \\ 
$-$0.35 & 0.00 & 5.00 & $-$0.40 & 0.97 \\ 
$-$0.35 & 0.00 & 10.00 & $-$0.29 & 0.97 \\ 
$-$0.35 & 0.00 & 12.00 & $-$0.25 & 0.97 \\ 
$-$0.29 & 0.40 & 2.50 & $-$0.45 & 1.03 \\ 
$-$0.29 & 0.40 & 5.00 & $-$0.37 & 1.06 \\ 
$-$0.29 & 0.40 & 10.00 & $-$0.27 & 1.07 \\ 
$-$0.29 & 0.40 & 12.00 & $-$0.24 & 1.06 \\ 
$-$0.25 & 0.00 & 2.50 & $-$0.46 & 0.98 \\ 
$-$0.25 & 0.00 & 5.00 & $-$0.38 & 0.99 \\ 
$-$0.25 & 0.00 & 10.00 & $-$0.27 & 1.00 \\ 
$-$0.25 & 0.00 & 12.00 & $-$0.23 & 0.99 \\ 
$-$0.09 & 0.40 & 2.50 & $-$0.42 & 1.09 \\ 
$-$0.09 & 0.40 & 5.00 & $-$0.34 & 1.12 \\ 
$-$0.09 & 0.40 & 10.00 & $-$0.25 & 1.13 \\ 
$-$0.09 & 0.40 & 12.00 & $-$0.22 & 1.13 \\ 
0.05 & 0.40 & 2.50 & $-$0.41 & 1.14 \\ 
0.05 & 0.40 & 5.00 & $-$0.34 & 1.18 \\ 
0.05 & 0.40 & 10.00 & $-$0.25 & 1.19 \\ 
0.05 & 0.40 & 12.00 & $-$0.22 & 1.20 \\ 
0.06 & 0.00 & 2.50 & $-$0.36 & 1.04 \\ 
0.06 & 0.00 & 5.00 & $-$0.28 & 1.07 \\ 
0.06 & 0.00 & 10.00 & $-$0.19 & 1.08 \\ 
0.06 & 0.00 & 12.00 & $-$0.16 & 1.08 \\ 
0.25 & 0.00 & 2.50 & $-$0.32 & 1.09 \\ 
0.25 & 0.00 & 5.00 & $-$0.24 & 1.12 \\ 
0.25 & 0.00 & 10.00 & $-$0.15 & 1.14 \\ 
0.25 & 0.00 & 12.00 & $-$0.12 & 1.14 \\ 
0.40 & 0.00 & 2.50 & $-$0.30 & 1.13 \\ 
0.40 & 0.00 & 5.00 & $-$0.22 & 1.16 \\ 
0.40 & 0.00 & 10.00 & $-$0.14 & 1.19 \\ 
0.40 & 0.00 & 12.00 & $-$0.11 & 1.19 \\ 
\hline
\end{tabular}
\caption{The mean magnitude and color of the red clump as predicted by the BaSTI isochrones for a range of population abundances and ages.}
\label{table:BaSTI} 
\end{table}

We also list the predictions of the updated BaSTI database$^{2}$ \citep{2018ApJ...856..125H} for
four metallicites. We downloaded isochrones with overshooting, diffusion, and mass-loss parameter $\eta =$ 0.30. This new version of the BaSTI model library represents a significant improvement with respect the previous release. We refer to \citet{2018ApJ...856..125H} for a detailed discussion on the various updates in both the input physics and numerical assumptions adopted for computing the new BaSTI stellar models. However, for the sake of completeness here we mention that version 2 of BaSTI models is based on: an updated solar heavy-elements distribution \citep{2011SoPh..268..255C}, updated nuclear reaction rates for both H- and He-burning phase (see \citealt{2010A&A...522A..76P} and references therein), revised outer boundary conditions, and updated conductive opacity evaluations \citep{2007ApJ...661.1094C} as well as an average mass loss efficiency during the RGB stage constrained on the basis of asteroseismic benchmarks \citep{2012MNRAS.419.2077M}. In particular, we note that the use of updated conductive opacity and a different mass loss efficiency directly impacts on the RC average color and magnitude, as will be shown in the following sections. Ideally, these isochrones would be used for all of the calculations, but their grid for the $\alpha-$enhanced heavy element distribution is not yet available. We also compute, for purposes of completeness, the predictions from the MIST\footnote{http://waps.cfa.harvard.edu/MIST/} version 1.2 isochrones without rotation\citep{2016ApJS..222....8D,2016ApJ...823..102C}, and the predictions from the PARSEC\footnote{http://stev.oapd.inaf.it/cgi-bin/cmd} v1.2S isochrones with the YBC option for bolometric corrections and mass loss parameter $\eta=0.2$ \citep{2012MNRAS.427..127B,2014MNRAS.444.2525C,2015MNRAS.452.1068C,2014MNRAS.445.4287T}. 

The different predictions are listed in Table \ref{table:BaSTIv2}. Relative to the [Fe/H] $\leq 0$ predictions from the BaSTI v1 isochrones, the MIST v1.2 and PARSEC v1.2S predictions for the red clump from are slightly fainter and bluer ($\approx$ 0.02 mag), whereas the BaSTI v2 predictions are approximately 0.04 mag bluer and 0.12 mag fainter. The results are summarized in Table \ref{table:BaSTIv2}. From the four sets of models, and for parameter values approximately appropriate for the SMC and LMC, the predictions are of a a $(V-I)_{RC,0}$ color that is approximately $0.02 \pm 0.02$ mag bluer, and of M$_{I,RC}$ that is $0.04 \pm 0.06$ mag fainter, than that of the BaSTI v1 isochrones.

\begin{table*}
\begin{center}
\begin{tabular}{| ll | cc| cc | cc | cc |} 
\hline
[Fe/H]  & $\tau$/Gyr & M$_{I,RC}$ & $(V-I)_{RC,0}$ & M$_{I,RC}$ & $(V-I)_{RC,0}$  & M$_{I,RC}$ & $(V-I)_{RC,0}$   & M$_{I,RC}$ & $(V-I)_{RC,0}$ \\
\hline
\multicolumn{2}{|c|}{Parameters} & \multicolumn{2}{c|}{BaSTI v1} & \multicolumn{2}{c|}{BaSTI v2}
 & \multicolumn{2}{c|}{MIST v1.2} & \multicolumn{2}{c|}{PARSEC v1.2S}  \\ 
\hline
$-$0.96 & 5.00 & $-$0.53 & 0.87 &  $-$0.44 & 0.84  &  $-$0.51 & 0.85  &  $-$0.54 & 0.84 \\ 
$-$0.96  & 10.00 & $-$0.36 & 0.85 & $-$0.18 & 0.79  & $-$0.34 & 0.83  & $-$0.38 & 0.83 \\ 
$-$0.66 & 5.00 & $-$0.48 & 0.92 &  $-$0.38 & 0.89  &  $-$0.46 & 0.90  &  $-$0.45 & 0.90 \\ 
$-$0.66  & 10.00 & $-$0.34 & 0.91 & $-$0.18 & 0.86  & $-$0.32 & 0.89  & $-$0.27 & 0.88 \\ 
$-$0.35 & 5.00 & $-$0.40 & 0.97 & $-$0.31 & 0.95  & $-$0.39 & 0.96  & $-$0.40 & 0.98 \\ 
$-$0.35  & 10.00 & $-$0.29 & 0.97 & $-$0.15 & 0.94  & $-$0.27 & 0.96  & $-$0.28 & 0.98 \\ 
0.06  & 5.00 & $-$0.28 & 1.07 & $-$0.17 & 1.05  & $-$0.27 & 1.07  & $-$0.26 & 1.11  \\ 
0.06  & 10.00 & $-$0.19 & 1.08 & $-$0.04 & 1.06 & $-$0.15 & 1.08  & $-$0.16 & 1.12 \\ 
0.25  & 5.00 & $-$0.24 & 1.12 & $-$0.11 & 1.11  & $-$0.18 & 1.14  & $-$0.21 & 1.17 \\ 
0.25  & 10.00 & $-$0.15 & 1.14 & 0.01 & 1.12  & $-$0.07 & 1.16  & $-$0.12 & 1.20 \\ 
\hline
\end{tabular}
\caption{The predictions for the absolute magnitude and color of the red clump as a function of age and metallicity from four different isochrone libraries: BaSTI v1 \citep{2004ApJ...612..168P,2006ApJ...642..797P,2007AJ....133..468C}, BaSTI v2 \citep{2018ApJ...856..125H}, MIST v1.2 \citep{2016ApJS..222....8D,2016ApJ...823..102C}, and PARSEC v1.2S \citep{2012MNRAS.427..127B,2014MNRAS.444.2525C,2015MNRAS.452.1068C,2014MNRAS.445.4287T}. A description of our application of these models can be found in Section \ref{subsec:Theory}.}
\label{table:BaSTIv2} 
\end{center}
\end{table*}

\subsection{Composite Population Effects for the Magellanic Cloud Red Clumps} \label{subsec:Composite}

The mean color and brightness of the red clump of a stellar population need not necessarily correspond to the color and brightness of a red clump at the mean age and metallicity of a stellar population, as is assumed by Equations \ref{EQ:VMI_Basti}, \ref{EQ:MI_Basti}, \ref{EQ:VMI_Basti2}, and \ref{EQ:MI_Basti2}. Among the contributing factors, stellar subpopulations with $0.50  \lessapprox \tau/\rm{Gyr}  \lessapprox 2.0$ produce proportionately more red clump stars (see Figure 1 of \citealt{2001MNRAS.323..109G} and associated discussion for more details). Thus, on average, the red clump should be younger and more metal rich than the total stellar population. 

To evaluate the amplitude of this correction for the LMC, we construct synthetic luminosity functions weighted by the star-formation history of \citet{2013MNRAS.431..364W}, and using the BaSTI v2 isochrones, due in part to their excellent online interpolation tools. This correction increases $(V-I)_{RC,0}$ by $\approx$0.03 mag, and increases $M_{I,RC}$ by $\approx$0.03 mag, in other words, the peak of the luminosity function becomes slightly redder and dimmer. The mean age of the stars in and near the red clump is between 3 and 4 Gyr, consistent with the arguments in \citet{2016ARA&A..54...95G} and in \citet{2018MNRAS.478.5017R}. Estimates for the composite population effects of the Small Magellanic Cloud are excluded from this work due to their greater uncertainty, but we expect them to be comparable to or slightly greater than those for the Large Magellanic Cloud. 
We note that the estimates mean ages for stars in the red clump region are slightly greater than those ($\tau \approx 2$ Gyr) assumed by \citet{2017ApJ...847..102Y} and \citet{2018ApJ...866...90C}. We consider this (arguably) small offset to be unsurprising, given the numerous remaining uncertainties. Were we to assume these ages, our derived color and magnitude for the LMC red clump would be slightly redder and brighter still.

\subsection{Empirical Measurements of the color of the Red Clump} \label{subsec:Empirical}

\begin{table*}
\centering
\begin{tabular}{|l | lllll | rlrl |}
\toprule
Population &  [Fe/H] & [$\alpha$/Fe] & $\tau$/Gyr &  $(m-M)_{V}$ & $E(V-I)$ & $(V-I)_{RC,0}$ & $M_{I,RC}$  & $(V-I)_{RC,\rm{BaSTI,v1}}$ & $M_{I,RC,\rm{BaSTI,v1}}$  \\ 
\hline
\multicolumn{1}{|l|}{} & \multicolumn{5}{c|}{Assumed Parameters} & \multicolumn{2}{l}{Derived Red Clump Centroid}
 & \multicolumn{2}{r|}{Predicted Red Clump Centroid}   \\ 
\midrule
47 Tuc (NGC 104)  & $-0.$75 & $+$0.26 & 12 & 13.37 & 0.045 &  0.88 & $-$0.18 &  0.90 & $-$0.34 \\ 
Large Magellanic Cloud & $-$0.64 & $+$0.03 & 6 & -- & -- & -- & -- &  0.95 &$-$0.39 \\
Small Magellanic Cloud & $-0.88$ & $-$0.02 & 5 & -- & -- & -- & -- & 0.87 &  $-$0.50 \\
Solar Neighborhood &  $-$0.03 & $+$0.03 & 4.7 & -- & -- & 1.02 & $-$0.23  & 1.03 & $-$0.36 \\
NGC 6791 & $+$0.34 & $+$0.04 & 8.3 & 13.51 & 0.174 & 1.15 & $-$0.08  & 1.17& $-$0.17 \\
\bottomrule
\end{tabular}
\caption{In the four right-most columns, we show both the best-fit measured intrinsic parameters of the red clump, and the BaSTI-predicted parameters of the red clump. The measured parameters assume the distance and reddening values shown in the 5th and 6th columns, and the predicted parameters assume the chemistry and age values shown in the 2nd, 3rd, and 4th columns. For our three comparisons, the BaSTI-predicted values are on average $0.02 \pm 0.01$ mag redder in $(V-I)$, and $0.13 \pm 0.03$ mag brighter in $M_{I}$. 
\\ All [Fe/H] and [$\alpha$/Fe] values are homogeneously derived by ASPCAP \citep{2016AJ....151..144G} from APOGEE spectra \citep{2017AJ....154...94M}. The distance for the solar neighborhood sample are taken from Gaia \citep{2016A&A...595A...1G} Data Release 2 \citep{2018A&A...616A...2L} with a parallax zero-point offset of  $\delta_{\pi}= -0.054$ mas derived by \citet{2019MNRAS.487.3568S}, and with the reddening values taken from \citet{2019ApJ...887...93G}. The star-formation histories of the SMC and LMC were derived by \citet{2013MNRAS.431..364W} from deep \textit{HST} imaging, and their distances were derived by the detached eclipsing binary studies of \citet{2014ApJ...780...59G} and \citet{2019Natur.567..200P}. The ages, distances, and reddening to 47 Tuc and NGC 6791 were respectively derived by  \citet{2020MNRAS.492.4254T} and \citet{2012MNRAS.419.2077M}, for which the detached eclipsing binaries were responsible for the distance estimates.  }
\label{table:pop_params}
\end{table*}

\begin{figure}
    \includegraphics[width=1.0\linewidth]{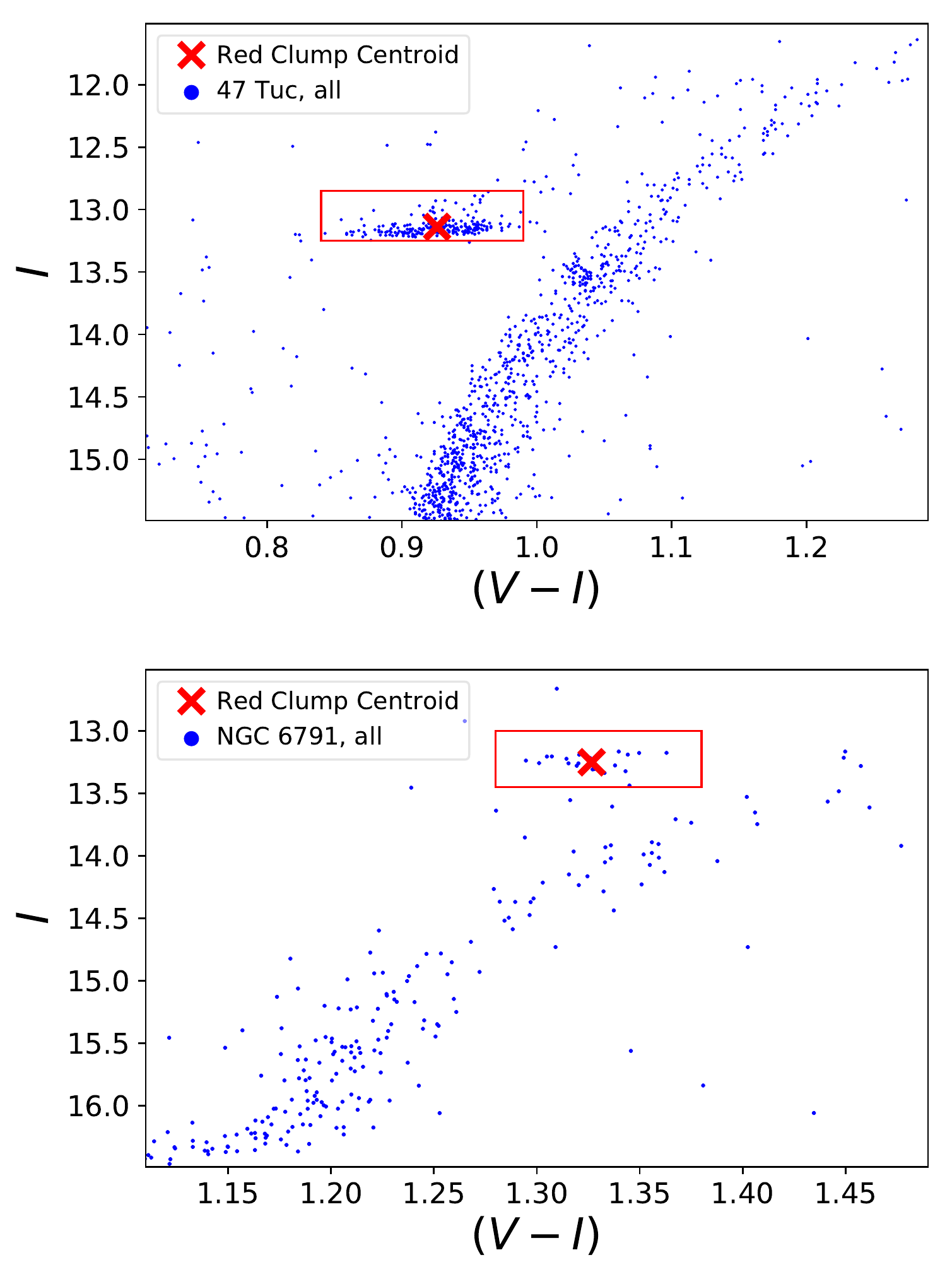}
    \caption{The color-magnitude diagrams of 47 Tuc (TOP) and NGC 6791 (BOTTOM). The stars selected as being part of the red clump are plotted in the red rectangles. \vspace{1em}}
    \label{figure:EmpiricalRC}
\end{figure}

In the prior section, we predicted the slope and zero-point of the color-metallicity relation of the red clump. In this subsection, we use the  measured color-magnitude diagrams of three stellar populations in the Milky Way to verify their accuracy. These are the globular cluster 47 Tuc (NGC 104), the solar neighbourhood stellar population, and the open cluster NGC 6791. These populations are selected as they host numerous red clump stars, and they are well-characterized in distance, reddening, and population parameters such as metallicity. They also provide a broad span in [Fe/H] which brackets the Magellanic Clouds. The color-magnitude diagrams of 47 Tuc and NGC 6791 are shown in Figure \ref{figure:EmpiricalRC}.

The population parameters that we adopt are derived from a range of methods and data sets. There is unfortunately no uniformly-derived set of population parameters available across these different populations. Thus we adopt the most tightly calibrated parameters available for the distances, reddening, metallicity, and age determinations (Table 2). 

\subsubsection{47 Tuc}

For 47 Tuc, we use the color-magnitude diagram of \citet{2009AJ....138.1455B}, and we restrict the sample to those stars with angular separation from the cluster center of $(\alpha,\delta)=(6.023625, -72.081276)$\footnote{Taken from \url{https://people.smp.uq.edu.au/HolgerBaumgardt/globular/orbits.html}, for details see \citet{2019MNRAS.482.5138B}.}, in arcseconds, of $400  \leq \delta \psi \leq 900 $, in order to reduce the effect of blending. We select the 241 red clump stars satisfying $0.84\leq (V-I) \leq 0.99$ and $12.85 \leq I \leq 13.25$, for mean apparent photometry of $(V-I)_{RC}=0.927 \pm 0.002$ and $I_{RC}=13.139 \pm 0.004$.  The value of this measured centroid is negligibly sensitive to the dimensions of the box, as shifting the box 0.02 mag in color or 0.03 mag in magnitude has a 0.001 mag effect on each of $(V-I)_{RC}$ and $I_{RC}$. We have verified that we obtain a nearly identical result, $(V-I)_{RC}=0.913 \pm 0.001$ and $I_{RC}=13.086 \pm 0.003$  , if we had instead used the color-magnitude diagram of \citet{2007AJ....133.1658S}, which is obtained from highly accurate and highly-precise Hubble Space Telescope observations in $F606W$ and $F814$ transformed into $V$ and $I$ using theoretical transformations. The slightly bluer and brighter horizontal branch is likely due to the gradient in helium enrichment in the cluster \citep{2011ApJ...736...94N,2012ApJ...744...58M}.

For the apparent distance, reddening to, and age of 47 Tuc, we adopt the values of $(m-M)_{V}=13.370 \pm 0.015$, $E(V-I)=0.045 \pm 0.01$, and $\tau=12.0$ Gyr from \citet{2020MNRAS.492.4254T}. These measurements are predominantly derived from spectroscopic and photometric measurements of two members of an eclipsing binary system. The value and error of the distance modulus are computed from the values of the four eclipsing binary members. That is somewhat unphysical, as the errors on those four individual distances are almost certainly correlated (or anti-correlated), but unfortunately both the sign and amplitude of that correlation are unknown. 

We estimate the population abundances from APOGEE spectra of 60 cluster stars, where we use the cluster selection criteria as \citet{2019AJ....158...14N}, we require $\log{g} \leq 3.0$, and we restrict the analysis to stars satisfying [N/Fe] $\leq +0.35$, a parameter space for which the consistency of the ASPCAP pipeline with other literature estimates has been  demonstrated \citep{2018AJ....156..126J}. We obtain the median abundances of [Fe/H]$=-0.75$ and [$\alpha$/Fe]$=+0.26$.

%Given that 82.5\% \citep{2017MNRAS.464.3636M} of the stars in 47 Tuc are helium-enriched by $\delta$Y=0.011 \citep{2018MNRAS.481.5098M}, the mean helium abundance is 0.008 higher. From the BaSTI database, we estimate that this should decrease $(V-I)_{RC}$ by 0.005, and decrease $M_{I,RC}$ by 0.024 mag. 

Our assumed population parameters for 47 Tuc are consistent with other literature values. From \citet{2014A&A...572A.108T}, \citet{2015AJ....149...71J}, and \citet{2016MNRAS.459..610M} we respectively find ([Fe/H],[Mg/Fe],[Ca/Fe])$=(-0.78,+0.44,+0.24)$, ([Fe/H],[Mg/Fe])$=(-0.73,+0.33)$, and [Fe/H]$=-0.68$. The distance ($\mu =$ 13.29, Michal Rozyczka and Aaron Dotter, private communication) is a little higher than the value of  $\mu = 13.23 \pm 0.02$  derived by  \citet{2019MNRAS.482.5138B} from Gaia DR2 data. \citet{2016A&A...590A..64S} estimated $\mu \approx 13.33$ by matching the observed red clump stellar distribution with models constructed from the same stellar evolution code used in this work, but with a different prescription concerning the mass loss efficiency during the red giant branch stage. Finally, \citet{2013Natur.500...51H} estimated an age of $\tau/\rm{Gyr} = 9.9 \pm 0.7$ from its white dwarf cooling sequence. \citet{2002A&A...395..481G} used stromgren photometry of the cluster to estimate $(m-M)_{V}=13.33$, $E(B-V)=0.04$, and an age of $\tau/\rm{Gyr} = 12$.

The stellar population of 47 Tuc is also characterized by a small spread in initial helium abundance Y \citep{2010MNRAS.408..999D,2011ApJ...736...94N,2017MNRAS.464.3636M,2018MNRAS.481.5098M}. Indeed, \citet{2013A&A...549A..41G} studied the spectra of 110 red clump stars in 47 Tuc and found the [Na/O] abundances -- and thus by proxy, the initial helium abundances -- were tightly anti-correlated with the $(B-V)_{0}$ colors. \citet{2016A&A...590A..64S} showed that the standard deviation in color and magnitude of the red clump of 47 Tuc is matched well by varying the initial helium abundance of the cluster's stars to be uniformly distributed in the interval $Y \in [0.256, 0.286]$. Using the BaSTI database, we find that this will shift the predicted red clump to being bluer and brighter, by $(\Delta(V-I)_{RC,0},\Delta M_{I,RC})=(-0.009,-0.042)$. We find the predicted properties of the red clump in 47 Tuc are thus $((V-I)_{RC,0},M_{I,RC})=(0.899,-0.336)$. In contrast, the observed values are $((V-I)_{RC,0},M_{I,RC})=(0.882,-0.186)$.

%Combining the offsets due to helium enrichment with the parameters from Table \ref{table:pop_params} satisfying $\tau = 12$ Gyr, [Fe/H] $\leq -0.30$, we find that the predicted 47 Tuc red clump has $(V-I)_{RC,0},M_{I,RC}=(0.90,-0.32)$. In contrast, the dereddened observed values are $0.87,-0.24$. 

%47 Tuc, VMI: [0.90733195 0.24274588 0.14145286]
%47 Tuc, MI: [-0.29635803  0.0982354   0.00255425]

\subsubsection{Solar Neighborhood}

For the solar neighbourhood, \citet{2019arXiv191000398C} have measured mean values of $M_{\rm{RC,J}} = -1.019$ and $M_{\rm{RC,Ks}} = -1.622$. However, the selection function that they used assumes the APOGEE red clump catalog \citep{2014ApJ...790..127B}, which incorporates spectroscopic priors. That may differ from the photometric definition of the red clump used for populations such as the LMC and SMC, which is an excess in the luminosity function of post-main-sequence stars at or very close to the location of the horizontal branch. 

\begin{figure}[ht]
 \includegraphics[width=1.0\linewidth]{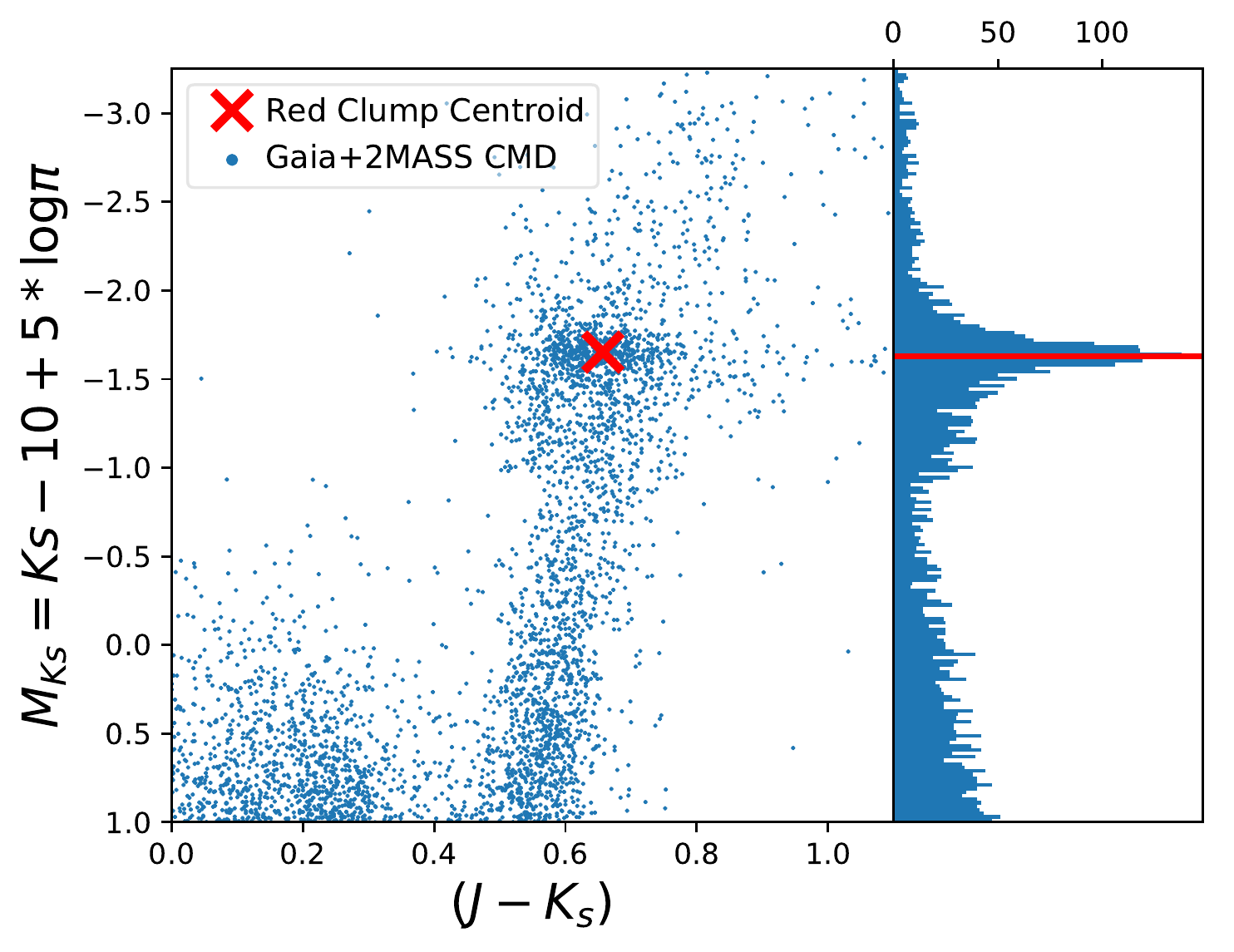}
 \caption{LEFT: The $(J-K_{s},M_{Ks})$ color-magnitude diagram of solar neighbourhood stars within $\sim 300$ pc, with $\delta \geq -30^{\circ}$. We plot every 5th data point for purposes of clarity. RIGHT: The $M_{Ks}$ luminosity function of the red giant branch stars, the red clump is measured to have a peak magnitude of $M_{Ks} = -1.63$.} 
\label{figure:GaiaRC}
\end{figure}

We produced a photometric selection of Red Clump stars in the solar neighborhood by the following steps.  We used the Gaia \citep{2016A&A...595A...1G} Data Release 2 \citep{2018A&A...616A...2L} solar neighbourhood measurements from the Gaia archive\footnote{https://gea.esac.esa.int/archive/}, where we used their cross-match to the 2MASS catalog  \citep{2006AJ....131.1163S}. We required $visibility\_periods\_used > 5$, $astrometric\_excess\_noise < 1$, and apply a zero-point offset to the parallaxes of  $\delta_{\pi}= -0.054$ mas  \citep{2019MNRAS.487.3568S}. We select nearby stars with precise parallax measurements, such that  $\pi \geq 3$ and $\pi / \sigma_{\pi} \geq 100$. 

 We  selected a low-reddening sample by requiring that the estimated reddening to the stars satisfies $E(g-r) \leq 0.04$ \citep{2019ApJ...887...93G}, for a mean reddening of $ \langle E(g-r) \rangle = 0.005$, or equivalently, $\langle E(J-K_{s}) \rangle = 0.003$ ( $R_{V} = 3.1$ \citealt{1999PASP..111...63F}). We use the fitting routine of \citet{2016MNRAS.456.2692N} to measure the properties of the red clump sample as an excess in the luminosity function over that of red giant stars, a method first developed by \citet{1997ApJ...477..163S} and \citet{1998ApJ...494L.219P}. This selects approximately 1500 red clump stars. We note that the stars used in our analysis are not themselves inputs into the 3D reddening maps of \citet{2019ApJ...887...93G}, as those maps are constructed from stars fainter than the saturation limits of Pan-STARRS 1 \citep{2016arXiv161205243F}.  

To link measurements in $(J,K_{s})$ to expected colors in $(V,I)$, we use the color-color relations from Table 1 of \citet{2016MNRAS.456.2692N} and linearly interpolate over a small interval, which are derived from the library of synthetic stellar colors of  \citet{2014MNRAS.444..392C}. 

%For stars satisfying $\pi \geq 3$ and $\pi / \sigma_{\pi} \geq 100$, we measure the red clump parameters using the same fitting routine as \citet{2016MNRAS.456.2692N}. For that sample, we measure $(J-K_{s},M_{\rm{RC,Ks}})=(0.657,-1.652)$. The derived parameters are not sensitive to the choice of cuts in $\pi/\sigma_{\pi}$ and $\pi$ in this regime of very precise values. The lack of sensitivity to $\pi$ suggests very little extinction to these stars, and indeed, from the 3D reddening maps of \citet{2019ApJ...887...93G} we find a mean reddening of $E(J-K_{s}),A_{Ks} \approx (0.015,0.01)$ to this sample, which we apply.

%There is however a dependence on the extinction $A_{G}$ \citep{2018A&A...616A...8A}. Fitting for $(J-K_{s}) vs A_{G}$ yields a reddening of $E(J-K_{s}) =0.04$ mag for this sample. In comparison, \citet{2005ApJ...619..931I} measured a local reddening density of $c_{J-K_{s}} \approx 0.20$ mag kpc$^{-1}$, which would yield 0.05 mag for this sample.  

Our best-fit values for the solar neighborhood red clump are thus
 $M_{\rm{RC,Ks}} = -1.63$, $\sigma_{M_{Ks}} = 0.14$, $(J-K_{s})=0.64$,  $M_{\rm{RC,I}} = -0.23$, and $(V-I)=1.02$, where the statistical errors are each less than 0.01 mag. The $((V-I)_{RC,0},M_{I,RC})$ compare well to the prior measurement of $M_{\rm{RC,I}} -0.28$, and $(V-I)=1.01$ from \citet{1998ApJ...494L.219P}, which was derived from Hipparcos data \citep{1997A&A...323L..49P}. The zero-point shift in the parallaxes of $\delta_{\pi}= -0.054$ mas  \citep{2019MNRAS.487.3568S} shifts $M_{\rm{RC,Ks}}$ by 0.03 mag. If we instead adopt the distances of \citet{2018AJ....156...58B}, which assume priors for Galactic geometry discussed in that work, a zero point offset to the Gaia parallaxes of $\delta_{\pi}= -0.029$ \citep{2018A&A...616A...2L}, we obtain the same color, but a slightly brighter red clump, with $M_{\rm{RC,Ks}} = -1.64$. We thus suggest a 0.01 mag error on $M_{Ks}$, due to this ambiguity as to which small correction we should apply to the DR2 parallaxes. 
 
 We also derive the following colors for the solar neighborhood red clump, by direct fitting to the Gaia and 2MASS data: $(G_{BP}-K_{s})=2.75$, $(G-K_{s})=2.18$, $(G_{RP}-K_{s})=1.52$, and $(H-K_{s})=0.15$.
 
 The color-magnitude diagram of solar neighbourhood stars selected to have low reddening, and the combined red giant and red clump luminosity function, is shown in Figure \ref{figure:GaiaRC}.
 
 %The solar neighbourhood color-magnitude diagram, and its K-band luminosity function, are shown in Figure \ref{figure:GaiaRC}.
 
 To compute the parameters for the solar neighborhood stellar populations, we star with the posterior age distributions for $\sim$20,000 APOGEE targets with Galactic positions satisfying $7 \leq R_{GC}/\rm{kpc} \leq 9$, and $|z| \leq 0.50 kpc$ that were computed by \citet{2019MNRAS.489.1742F}. We then used the Galactocentric positions derived by \citet{2020A&A...638A..76Q} to re-weigh this sample to have the same mean separation from the Galactic plane ($\sim$52 pc) and same standard deviation thereof (123 pc) our photometric sample of red clump stars. $(\langle \rm{[Fe/H]} \rangle, \langle [\alpha/\rm{Fe}] \rangle,\langle \tau/\rm{Gyr} \rangle) =(-0.03,+0.03,4.7)$. The composite population corrections (as in Section \ref{subsec:Composite}) decreases $(V-I)_{RC,0}$, and decreases $M_{I,RC}$ by $\approx 0.03$ mag, in other words, the peak of the luminosity function becomes slightly bluer and brighter. 
 
  These are remarkably similar to the values of $(\langle \rm{[Fe/H]} \rangle, \langle [\alpha/\rm{Fe}] \rangle,\langle \tau/\rm{Gyr} \rangle) =(-0.04,+0.03,5.1)$ that we derive by applying the same method to GALAH survey data
 \citep{2015MNRAS.449.2604D,2019A&A...624A..19B}.

 %We thus derive $(\langle \rm{[Fe/H]} \rangle, \langle [\alpha/\rm{Fe}] \rangle,\langle \tau/\rm{Gyr} \rangle) =(-0.06,+0.04,5.1)$
 
 %The solar neighbourhood population parameters that we assume, $(\langle \rm{[Fe/H]} \rangle, \langle [\alpha/\rm{Fe}] \rangle,\langle \tau/\rm{Gyr} \rangle) =(-0.06,+0.04,5.1)$ are from \citet{2019MNRAS.489.1742F}, who computed posterior age distributions for $\sim$20,000 APOGEE targets with Galactic positions satisfying $7 \leq R_{GC}/\rm{kpc} \leq 9$, and $|z| \leq 0.50 kpc$. Using the Galactocentric positions derived by \citet{2020A&A...638A..76Q}, we re-weighed the sample of \citet{2019MNRAS.489.1742F} to have the same mean separation from the Galactic plane ($\sim$52 pc) and same standard deviation thereof (123 pc) our photometric sample of red clump stars. 
 
 %These are remarkably similar to the values of $(\langle \rm{[Fe/H]} \rangle, \langle [\alpha/\rm{Fe}] \rangle,\langle \tau/\rm{Gyr} \rangle) =(-0.06,+0.03,5.1)$ derived by \citet{2019A&A...624A..19B} from GALAH survey data
 %\citep{2015MNRAS.449.2604D}.
 
 The predicted solar neighbourhood red clump color and magnitude, assuming the parameters from \citet{2019MNRAS.489.1742F}, fitting over the full span of Table \ref{table:BaSTI} and applying the corrections for composite population effects, are $((V-I)_{RC,0},M_{I,RC})=(1.03,-0.36)$. In contrast, the observed values are $((V-I)_{RC,0},M_{I,RC})=(1.02,-0.25)$.

 \subsubsection{NGC 6791}
 
For NGC 6791, we use the color-magnitude diagram of \citet{2003PASP..115..413S}, which was updated for the work of \citet{2012A&A...543A.106B} and now includes corrections for differential reddening. We limit the selection to stars with Gaia-derived proper motions that are within 1.25 mas of the cluster central value of $\mu_{\alpha}$ cos$\delta$, $\mu_{\delta}$ = ($-$0.42,$-$2.28) reported by \citet{2018AJ....156..142D}. We select 24 red clump stars that have $1.24 \leq (V-I) \leq 1.38$ and $13.45 \leq I \leq 13.00$, for mean apparent photometry of $( (V-I)_{RC},\,I_{RC})_{\rm{NGC\,\,6791}} = (1.327 \pm 0.003,\,13.252 \pm 0.013)$. The value of this measured centroid is negligibly sensitive to the dimensions of the box, as shifting the box to remove the bluest point, which does appear as the most uncertain point, reduces $(V-I)_{RC}$  by 0.002 mag and increases  $I_{RC}$ by 0.003 mag effect.    The color-magnitude diagram of NGC 6791 is shown in the bottom panel of Figure \ref{figure:EmpiricalRC}. 

The analysis of \citet{2012A&A...543A.106B}, which incorporates information from the masses and radii of eclipsing binaries as well as the morphology of the three-bandpass color-magnitude diagram, yields the cluster parameters of  $(m-M)_{V}=13.51 \pm 0.06$, $E(V-I)=0.17 \pm 0.02$, and $\tau=8.3$ Gyr. The value and error of the distance modulus is computed from the values of the four eclipsing binaries listed in Table 1 of \citet{2012A&A...543A.106B}. That is somewhat unphysical, as the errors on those four individual distances are almost certainly correlated (or anti-correlated), but unfortunately both of the sign and amplitude of that correlation are not available.  From APOGEE spectra, the median abundances of 55 cluster stars with surface gravity $\log{g} \leq 3.0$ are [Fe/H]$=+0.34$, [$\alpha$/Fe]$=+0.04$. 

These adopted population parameters are consistent with other literature values. \citet{2015ApJ...799..202B} measured $(\langle \rm{[Fe/H]} \rangle, \langle 1/5[O+Mg+Si+Ti+Ca/\rm{Fe}] \rangle) =(+0.30,-0.02)$, and \citet{2018ApJ...867...34V} measured $(\langle \rm{[Fe/H]} \rangle, \langle 1/4[Mg+Si+Ti+Ca/\rm{Fe}] \rangle) =(+0.31,+0.06)$. \citet{2010Natur.465..194G} estimated an age of 8 Gyr from analysis of the cluster's white dwarf cooling sequence, and \citet{2019ApJ...874..180M} estimated an age of $\tau/\rm{Gyr} = 8.2$ from an asteroseismic analysis of the cluster's giants. One discrepant result is that of \citet{2015ApJ...811...46A}, who applied the method of main-sequence fitting to  $BVI_{c}JHK_{s}$ photometric data of the cluster, and derived best-fit values of [M/H]=$+0.42$, $(m-M)_{V}=13.38$, $E(V-I)=0.137$, and $\tau/\rm{Gyr} = 9.5$. \citet{2007AJ....133.1585A} used stromgren photometry of the cluster to [Fe/H]$=+0.45$ $(m-M)_{V}=13.60$, $E(B-V)=0.155$, and an age of $\tau/\rm{Gyr} = 7$.

Using predicted parameters from Table \ref{table:pop_params} satisfying $5 \leq \tau/\rm{Gyr} \leq 10$ Gyr, [Fe/H] $\geq  0$, we find that the predicted NGC 6791 red clump has $((V-I)_{RC,0},M_{I,RC}=(1.17,-0.17))$. In contrast, the dereddened observed values are $((V-I)_{RC,0},M_{I,RC}=(1.15,-0.09)$.

\subsection{A Combined Theoretical and Empirical Prediction for the Red Clumps of the Magellanic Clouds}

The offsets between theory and data for the three stellar populations used in this work are highly consistent which supports the use of empirical re-calibration of the models. For 47 Tuc, the solar neighborhood, and NGC 6791, we measure red clumps that are respectively 0.02, 0.03, and 0.02 mag bluer in $(V-I)$, and 0.16, 0.09, and 0.09 mag fainter in $M_{I}$, relative to the model predictions, as listed in Table \ref{table:BaSTI} and Equations (\ref{EQ:VMI_Basti}), (\ref{EQ:MI_Basti}), (\ref{EQ:VMI_Basti2}), and (\ref{EQ:MI_Basti2}). The offset between the data and the old BaSTI models is nearly equal to the offset between the old BaSTI models and the updated BaSTI models, where the latter are tyypically $\sim$0.03 mag bluer and $\sim$0.13 mag fainter.

Assuming these mean offsets of 0.02 mag and 0.13 mag, we derive predicted colors and magnitudes for the Magellanic Clouds of $((V-I)_{RC,0,\rm{LMC}},M_{RC,I,\rm{LMC}}) = (0.93,-0.26)$ and $((V-I)_{RC,0,\rm{SMC}},M_{RC,I,\rm{SMC}}) = (0.85,-0.25)$.

\section{Verifying the Consistency of Our Absolute Magnitude and Intrinsic Color Calibrations} \label{sec:Consistency}

The independent calibrations of the absolute magnitude and intrinsic color of the red clump enable us to verify that our estimates are consistent with one another. By definition,
\begin{equation}
    \mu = I_{RC} - M_{I,RC} - A_{I},
\end{equation}
where $I_{RC}$ is the apparent magnitude of the red clump, and $A_{I}$ is the extinction in $I$-band along a sightline. If we assume a standard extinction curve, then, $A_{I} = 1.31E(V-I)$ ($R_{V}=3.1$, \citealt{1999PASP..111...63F}), and therefore:
\begin{equation}
   \mu = I_{RC} - M_{I,RC} - 1.31((V-I)_{RC}-(V-I)_{RC,0}).
\end{equation}

We downloaded the OGLE-III photometry for the LMC field 100.1 \citep{2008AcA....58...89U}. This sightline was selected to be towards the center of the LMC, enabling us to compare our result to the measured distance for that galaxy, without needing to correct for a geometric tilt. 

We derive $\mu_{\rm{LMC}} = 18.45 \pm 0.02\,(\rm{stat}) \pm 0.04\,(\rm{sys})$, which is consistent with the value of  $\mu_{\rm{LMC}}=18.48$ measured from the analysis of twenty eclipsing binary systems \citep{2019Natur.567..200P}. Among the potential systematic errors, a hypothetical uncertainty of 0.3 in $R_{V}$, of 0.01 in $(V-I)_{RC,0}$, and of 0.04 in $M_{I,RC}$ respectively propagate as uncertainties of 0.01, 0.01, and 0.04 in the derived value of $\mu_{\rm{LMC}}$, for a combined systematic uncertainty of of 0.04 mag.

\begin{figure}
 \includegraphics[width=1.00\linewidth]{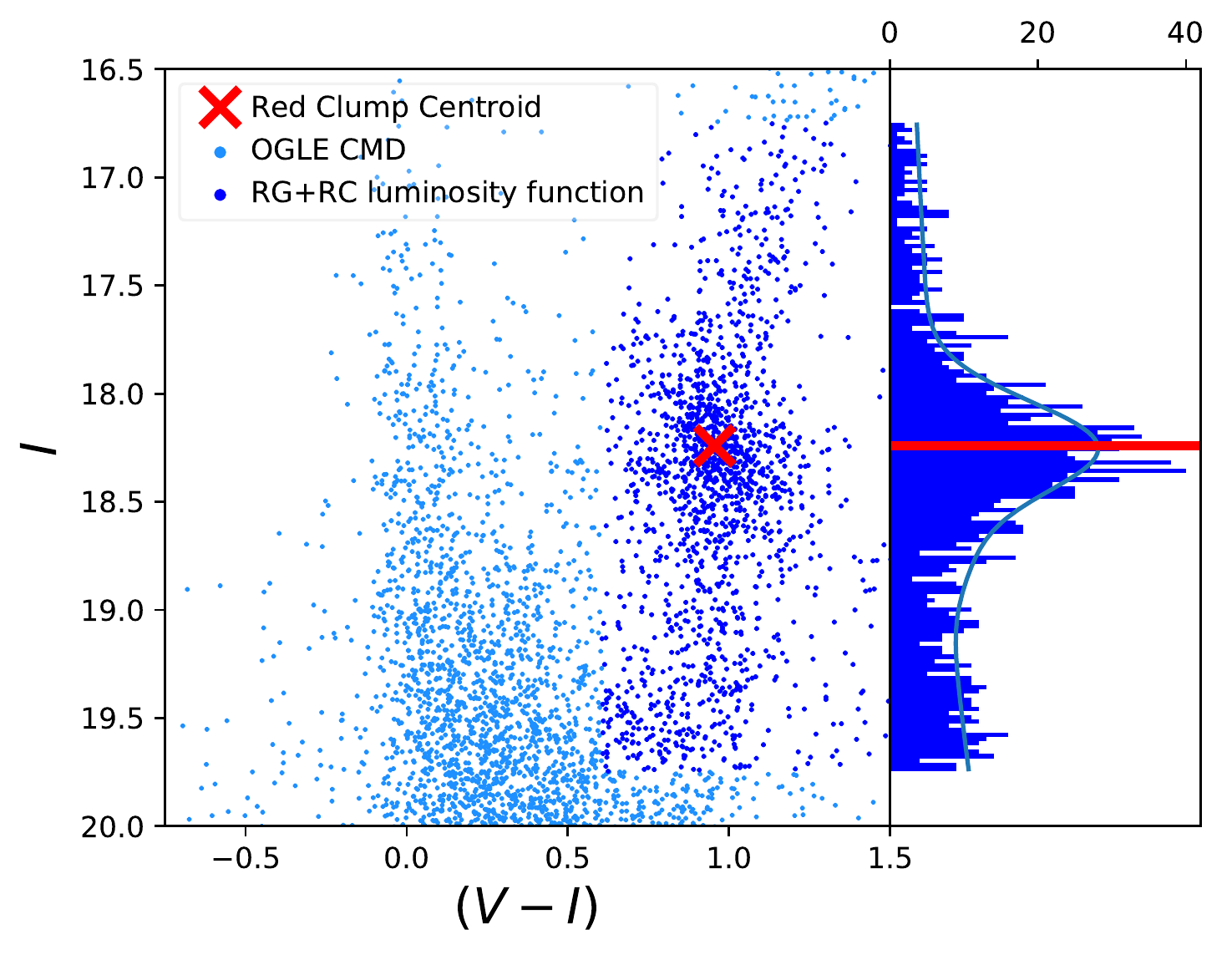}
 \caption{OGLE-III Color-magnitude diagram and luminosity function of part of the field lmc100.1. The combination of intrinsic red clump properties estimated in this work, with its apparent properties estimated from the color-magnitude diagram, yield a distance modulus of  $\mu_{\rm{LMC}} = 18.45 \pm 0.02\,(\rm{stat}) \pm 0.04\,(\rm{sys})$, consistent with the standard value of $\mu_{\rm{LMC}}=18.48$ \citep{2019Natur.567..200P}. 
 } 
\label{figure:OGLE100}
\end{figure}

This method, when applied to the SMC photometry \citep{2008AcA....58..329U}, did not yield as unambiguous a test as it did for the LMC. The red clump of the SMC appears to have a substantially more skew-symmetric magnitude distribution than that of the LMC. If we use the mean of the magnitudes of the red clump, we derive $\mu_{\rm{SMC}} \approx$ 18.76, whereas if we use the mode we derive $\mu_{\rm{SMC}} \approx$ 18.94. It is only the latter that is consistent with the standard value of $\mu_{\rm{SMC}} = 18.95 \pm 0.07$, that is measured from the analysis of four eclipsing binary systems \citet{2014ApJ...780...59G}. An assessment of consistency in this case will thus require a much more detailed analysis of the color-magnitude diagram of the SMC. 

%We note that there is some history behind the work done in this Section. Some of the first distances estimates to the Magellanic Clouds derived from photometry of red clump stars yielded a ``short distance scale", with $\mu_{\rm{LMC}}=$18.08 and $\mu_{SMC}=$18.56 ( \citealt{1998AcA....48....1U}), and separately, $\mu_{\rm{LMC}}=$18.36 and $\mu_{SMC}=$18.82 \citep{1998ApJ...500L.137C}. The reason for this discrepancy is partially the requirement to account for population effects (metallicity, age, etc) on the optical properties of the red clump \citep{2001MNRAS.323..109G,2003MNRAS.343..539P}, and partially that it was not clear how to account for these population effects. This was confirmed with the observations of the red giant branches of 15 local group dwarfs by \cite{2010AJ....140.1038P}. 

%, and for the SMC fields 100.1 and 100.2 \citep{2008AcA....58..329U}. These sightlines are selected to be towards the centers of those galaxies, enabling us to assume the measured distances for these galaxies, without needing to correct for uncertain geometric tilts. 

\section{The Reddening Toward the Magellanic Clouds} \label{sec:Analysis}

The Magellanic Clouds are among the best-studied sightlines in astronomy, and thus we have multiple options available to use independent tracers to test the available reddening maps derived from different choices for the color of the red clump. In this investigation, we use newly available observations of the colors of LMC Cepheids observed with \textit{HST}, classical Cepheids observed from the ground, and RR Lyrae stars of the ab-type observed from the ground. 

\subsection{Calibration with Cepheids} \label{subsec:Cepheids}

The intrinsic color of Cepheids can be estimated by means of well-calibrated period-color relations. They can then be treated as their own class of standard crayons, and thus the difference between their observed and intrinsic colors is an estimate of reddening. By comparing that estimate in the mean to the observed colors of red clump stars, we can estimate the intrinsic color of the latter.  

We use two separate sets of data for this comparison. The first is that of 70 LMC Cepheids observed by \citet{2019ApJ...876...85R}. These Cepheids have been observed in the WFC3-UVIS optical bands $F555W$ and $F814W$, in two HST programs: GO-14648 and GO-15146 (PI: Riess) across the inner part of the LMC.  The color $(F555W-F814W)$ is similar to the $(V-I)$ color studied here to evaluate LMC reddening, We exclude two Cepheids, OGLE1940 and OGLE1945 with colors of 1.3 and 1.6 mag, 5 and 8 $\sigma$ from the mean and indicative of high circumstellar reddening which is distinct from the interstellar reddening measured by the maps. The advantage of this data is that {\it HST} provides unmatched resolution and stability to overcome errors in Cepheid photometry which reside in some ground-based data due to crowding and zeropoint variations in the LMC.

The second dataset is that of 723 Cepheids from the catalogue of \citet{2015AJ....149..117M}. That investigation used a combination of new infra-red observations and optical $VI$ photometry from the OGLE survey \citep{2008AcA....58..163S,2013AcA....63..159U}. The actual catalog of \citet{2015AJ....149..117M} is larger than this sample -- we restricted our analysis to Cepheids that were used in the final fit of that investigation, and which were classified as fundamental rather than first-overtone pulsators. The advantages of this catalog relative to the {\it HST} one are that it is $\sim 10 \times$ more numerous, and that it spans a $\sim 2 \times$ larger area, including sightlines with slightly higher reddening.

We adopt two independent benchmarks for the unreddened Cepheid $(F555W-F814W)$ period-color relations. The first is empirical, from \citet{2006ApJ...652.1133M}. They used \textit{HST} data to study Cepheids in the nearby maser-host galaxy NGC 4258. Their sample includes 63 Cepheids with periods from 3 to 44 days toward an outer region, $\sim$ 20 Kpc from the center of that face-on galaxy where the host reddening should be minimal. The mean metallicity of that region is similar to that of the LMC \citep{2016ApJ...826...56R}, as the outer region of NGC 4258 was shown to have a metallicity $\sim$ 0.3 dex below solar \citep{2016ApJ...830...10H}, based on measurements of the $R23$ parameter in HII regions and presented on the \citet{1994ApJ...420...87Z} scale. The foreground reddening is also very low, $E(V-I)=0.022$, leaving little residual uncertainty after its removal. We note that the inclination angle toward this disk is $i = (86.93 \pm 0.22)^{\circ}$ \citep{2013ApJ...775...13H}, so nearly face-on.  We computed a weighted fit for a linear period-color relation to this data which yields $(F555W-F814W)_{0}= (0.895 \pm 0.012) + (0.317 \pm 0.038) (\log [ P/\rm{days} ] - 1)$.

%0.319 0.895
%[[ 1.38352231e-03 -9.94797946e-05]
% [-9.94797946e-05  1.35043007e-04]]
%[0.03719573 0.0116208 ]

The second unreddened benchmark is from the work of \citet{2013MNRAS.434.2866F}, who computed theoretical period-color relations from stellar pulsational models in the WFC3 $F555W-F814W$ system. Their predictions for various compositions can be found in their Table 1. We use the entry from that Table that they recommend for the LMC, which is computed from stellar models with initial metals abundance $Z=0.008$ and initial helium abundance $Y=0.25$ and yields 
$(F555W-F814W)_{0}=0.90+0.31(\log [ P/\rm{days} - 1] )$. That is consistent with that of the empirical relation within the errors of the latter.

\begin{figure*}
 \includegraphics[width=1.00\linewidth]{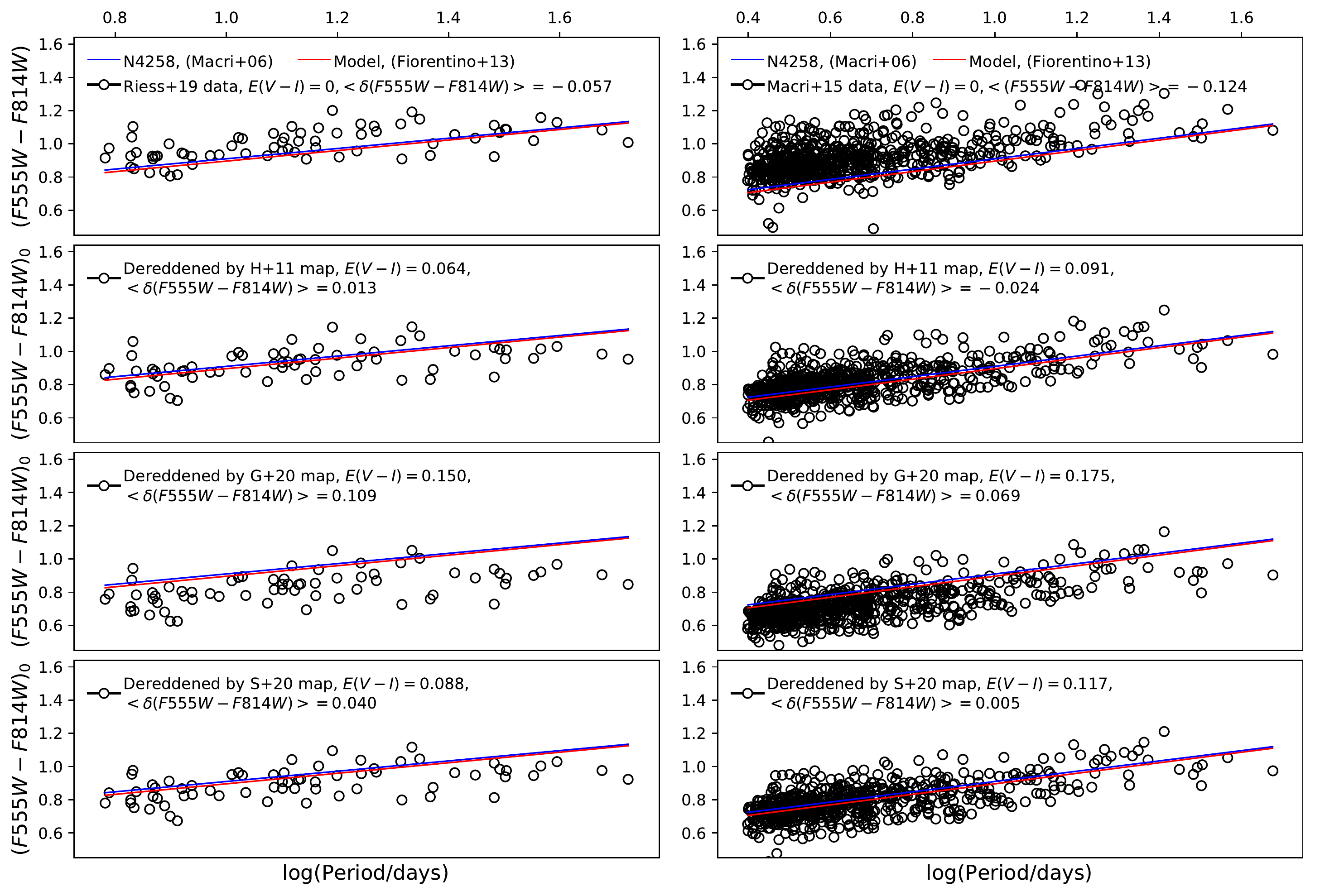}
 \caption{Comparison between the {\it HST} data (left column, \citealt{2019ApJ...876...85R}) and ground-based data (right column, \citealt{2015AJ....149..117M}), with predicted period-color relations, under the assumption of no reddening (top tow), and three different reddening maps (bottom 3 rows). In the legends, the value of $< {\delta}(F555W-F814W) >$ shows the mean excess of the \citet{2013MNRAS.434.2866F} relation relative to that of the observations. In other words, a positive value suggests that the Cepheids have been too de-reddened.}
\label{figure:CepheidWFC3_PeriodColorTotal}
\end{figure*}

In Figure \ref{figure:CepheidWFC3_PeriodColorTotal}, we plot the LMC Cepheid color $(F555W-F814W)$ as observed vs $\log{\rm{Period/days}}$ in the top panels. Then, in panels 2, 3, and 4 we show the derived dereddened colors of the Cepheids after using the reddening maps of \citet{2011AJ....141..158H}, \citet{2020ApJ...889..179G}, and \citet{2020arXiv200602448S}. The left panels show the  {\it HST} data \citep{2019ApJ...876...85R}, and the right panels show the scatter of the ground-based data \citep{2015AJ....149..117M}. In each of the three panels, the reddening-free empirical period-color relation \citep{2006ApJ...652.1133M}, is shown as the blue line, and the theoretical period-color relation \citep{2013MNRAS.434.2866F} is shown as the red line which are quite similar.

Without dereddening, the observed period-color distribution lies at redder colors than the theoretical prediction of \citet{2013MNRAS.434.2866F}, by a mean of 0.054 mag  in $(F555W-F814W)$ the {\it HST} data, which approximately corresponds to 0.048 mag in $E(V-I)$, assuming $R_{V}=3.1$ \citep{1999PASP..111...63F}. That is not surprising -- the reddening toward the LMC is definitely greater than 0. 

The application of reddening derived from Cepheids to the three red clump colors maps of \citet{2011AJ....141..158H}, \citet{2020ApJ...889..179G}, and \citet{2020arXiv200602448S} provide three pairs of estimates of $(V-I)_{RC,0}$ for each of the space-based and ground-based data. The three estimates will differ as the three maps have slightly different methodologies to measure the red clump color and slightly different resolutions. We take the mean of the three measurements as our derived value of $(V-I)_{RC,0}$. The error on  $(V-I)_{RC,0}$ is qaudratic sum of the scatter in those three values, and of the statistical error in the mean differences due to the scatter in the colors of the Cepheids. For the ground-based Cepheid data, we also incorporate a $\sim$0.02 mag error in the conversion from $VI$ photometry to {\it HST}-UVIS photometry, identified by \citet{2019ApJ...876...85R}. 

Assuming the \citet{2013MNRAS.434.2866F} relation, we derive $(V-I)_{RC,0,\rm{LMC}} = 0.93 \pm 0.02$ from the {\it HST} data and $(V-I)_{RC,0,\rm{LMC}} = 0.90 \pm 0.02$ from the ground-based data. If we assume the relation derived from the work of \citet{2006ApJ...652.1133M}, we derive $(V-I)_{RC,0,\rm{LMC}} = 0.94 \pm 0.03$ from the {\it HST} data and $(V-I)_{RC,0,\rm{LMC}} = 0.91 \pm 0.04$ from the ground-based data.

We comment on the possible origins of these differences. The first discrepancy is that the results derived from the relation of \citet{2006ApJ...652.1133M} are $\sim$0.01 mag redder than the results derived from the \citet{2013MNRAS.434.2866F} relation. That can be explained if, for example, the internal reddening of the outer field of NGC 4258 is not zero as we have assumed here, but $\langle E(V-I) \rangle = 0.01$. 

The second discrepancy is that the \textit{HST} data lead to derived values of $(V-I)_{RC,0}$ that are $\sim$0.03 mag redder than those from the OGLE (ground-based) data. A partial contributor to this discrepancy has been identified by \citet{2019ApJ...876...85R}: there is a 0.02 mag mean offset between directly-measured space-based colors, and space-based colors derived from ground-based colors. It is not clear if that offset is due to errors in the zero-points, errors in the assumed filter transmission curves, or greater blending in the ground-based data. Separately, a granular analysis of the data suggests that 0.01 mag of the offset may be due to a statistical fluke. We find that dereddening the data using the maps of \citet{2020arXiv200602448S}, with the analysis restricted to the period range that the two Cepheid samples have in common $(\log[ P/\rm{days}] \geq 0.75 )$, yields a 0.01 mag offset in color if we use the same ground-based photometric colors for both. 

%The first is that the result from the space-based data is $\sim$0.03 mag redder than that from the ground-based data. That is likely predominantly be due to a fluke, as is consistent with our errors. If we restrict the ground-based analysis to those 60 Cepheids that are common to both samples, the results from the ground-based data shift by 0.03 mag.

Given these issues, we adopt the mean of the estimates from the ground-based sample and space-based samples, and assume the theoretical period-color relation as primary, for  $(V-I)_{RC,0,\rm{LMC}} = 0.914 \pm 0.02 \approx 0.91 \pm 0.02$.

\subsection{Calibration with RR Lyrae (ab-type)} \label{subsec:RRab}

RR Lyrae stars are horizontal branch stars whose location in the Hertzsprung-Russel diagram places them in the instability strip. The relationship between their metallicities, pulsational periods, and $VIJHK_{s}$ magnitudes and colors are predicted by stellar theory \citep{2004ApJS..154..633C}. There are many types of RR Lyrae stars, in this work we use RR Lyrae of the ab-type (here after denoted ``RRab").  Population effects for RRab should not be large, as they probe a narrow region in the age-helium-metallicity  phase space of stellar populations, typically only old ($\tau \gtrapprox$ 10 Gyr) and metal-poor ([Fe/H] $\lessapprox -1.0$) stars  \citep{1994ApJ...423..248L,2004A&A...426..641C}.

\begin{table*}
\centering
\begin{tabular}{|l|cc|cc|}
\toprule
Method  & $(V-I)_{\rm{RC,0,LMC}}$ & $\langle \delta \psi \rangle$ & $(V-I)_{\rm{RC,0,SMC}}$ & $\langle \delta \psi \rangle$ \\
\hline
BaSTI v1 isochrones no empirically-derived offsets & $\approx$0.92 & 2.8  & $\approx$0.87 & 1.6  \\
BaSTI v1 isochrones with empirically-derived offsets  & $\approx$0.90 & 2.8  & $\approx$0.85 & 1.6 \\
BaSTI v1 '' plus composite population effects  & $\approx$0.93 & 2.8  & -- & 1.6 \\
\textit{HST}-observed Cepheids assuming an empirical color-period relation & $0.94 \pm 0.03$ & 1.4  & --  & --  \\
\textit{HST}-observed  Cepheids assuming an theoretical color-period relation  & $0.93 \pm 0.02$ & 1.4   & --   & --  \\
OGLE-observed Cepheids assuming an empirical color-period relation & $0.91 \pm 0.04$ & 1.2  & --   & --  \\
OGLE-observed Cepheids assuming an theoretical color-period relation  & $0.90 \pm 0.02$ & 1.2  & --   & --   \\
RRab assuming an empirical period-color relation  & $0.89 \pm 0.02$  & 3.0   & $0.84 \pm 0.02$  & 1.4   \\
\bottomrule
\end{tabular}
\caption{Summary of the estimates for the intrinsic colors of the LMC and SMC red clumps from the different methods explored in this work. The variable $\langle \delta \psi \rangle$  denotes the mean angular separation in degrees between the center of the host galaxy and the coordinates of the sources contributing to the measurement. In this work, we assume LMC and SMC centers of ($\alpha$,$\delta$)=(05:23:35,$-$69:45:22) and (00:52:38,$-$72:48:01).}
\label{table:methods}
\end{table*}

An RRab star's intrinsic color can be estimated from its pulsational properties and metallicity, which can be measured independently of distance and reddening. We use two different relationship. The first is the theory-derived, widely-used (e.g.  of \citealt{2015ApJ...811..113P,2016ApJ...821L..25K}) relationship of \citet{2004ApJS..154..633C}:
\begin{equation}
\begin{split}
(V-I)_{\rm{RRab},0} = & 0.55 + 1.13 \log \bigl( \rm{P}/(0.60\,\rm{days}) \bigl) \\ 
& + 0.070 \bigl(\rm{[Fe/H]}+1.00\bigl) + 0.11\bigl(\rm{[Fe/H]}+1.00\bigl)^{2},
\end{split}
\label{EQ:Catelan}
\end{equation}
where the models used by \citet{2004ApJS..154..633C} assume that $\log{Z} = (\rm{[Fe/H]}-1.765)$. We do not expect a perfect match with this model, as it reports synthetic colours for the RR Lyrae under the assumption that they are static stars, whereas the observational information that is available is that of $\langle V \rangle$ and $\langle I \rangle$, which are intensity-weighted mean magnitudes. We could not find predictions for the offset between $(V-I)_{\rm{Static}}$ and $\langle V \rangle - \langle I \rangle$, but the classic work of \citet{1995ApJS...99..263B} suggests that the synthetic $(B-V)_{s}$ color should be between 0.00 and 0.07 mag redder than the difference between the intensity-weighted magnitudes, $\langle B \rangle$ and $\langle V \rangle$.

The second relationship is empirical, which we derive from the compilation of \citet{2017AJ....153...96M}, using reddening corrections from \citet{2019MNRAS.490.4254N}, restricting the sample to those stars satisfying $ E(V-I) = 1.40 E(B-V) \leq 0.10$ ($R_{V}=3.1$, \citealt{1999PASP..111...63F}):
\begin{equation}
\begin{split}
    (\langle V \rangle -\langle I \rangle)_{\rm{RRab},0} = & 0.49 +  0.69 \log \bigl( \rm{P}/(0.60\,\rm{days}) \bigl) \\ & + 0.043 \bigl(\rm{[Fe/H]}+1.00\bigl),
    \end{split}
    \label{EQ:Monson}
\end{equation}

where the metallicities are taken from the work of \citet{1997A&AS..125..313F}, and the colors are the differences in the intensity-mean magnitudes. The reddening cut of $ E(V-I) = 1.40 E(B-V) \leq 0.10$ is implemented as the higher-reddening values are almost certainly over-estimates. The correlation between de-reddened color and reddening, if all RRab are included, is given by $\rho=-0.36$ with a p-value of 0.03.

The calibration provided by Equation \ref{EQ:Monson} enable us to use the RRab stars in the LMC and SMC as standard crayons. We can compare the $E(V-I)$ derived from RRab stars to the $(V-I)_{RC}$ measured by \citet{2020ApJ...889..179G} and \citet{2020arXiv200602448S} to calibrate the zero point for the latter. %When using the map of \citet{2020arXiv200602448S}, we restrict our analysis to that of the inner LMC and SMC, where the resolution of the maps is less than 4{\arcsec}. 

The LMC and SMC RR Lyrae are taken from the catalog of \citet{2016AcA....66..131S}, from which we used the subsample analyzed by \citet{2016AcA....66..269S}, which is selected to have more secure lightcurves and more secure metallicities derived from the Fourier components of the lightcurves. That sample totals 24,133 LMC and SMC RRab stars with OGLE-IV lightcurves. We use the metallicities that are on the \citet{1995AcA....45..653J} scale, for which \citet{2016AcA....66..269S} estimates the typical error to be 0.10 dex. We then restrict the analysis to RRab stars with $(V-I)_{\rm{Observed}} \geq 0.35$, $-1.75 \leq \rm{[Fe/H]}_{\rm{Fourier}} \leq -0.25$, and $0.45 \leq P/{\rm{days}} \leq 0.80$. That leaves a sample of 16333 LMC RRab stars, and 1612 SMC RRab stars. 

We compare the RRab-derived reddening maps to the RC-derived maps as follows. We restrict the comparison to the grid elements in the maps of \citet{2020ApJ...889..179G} that satisfy $\sigma_{V-I,RC} \leq 0.06$ and $E(V-I)_{RC} \leq 0.40$ to reduce the confounding effects of high reddening and differential reddening. The mean separation to the centers of the LMC and SMC are 2.6 and 1.6 degrees respectively. Similarly, for the grid of \citet{2020arXiv200602448S}, we require $\sigma_{V-I,RC} \leq 0.05$ and $E(V-I)_{RC} \leq 0.40$. The mean separation to the centers of the LMC and SMC are 3.4 and 1.2 degrees respectively. Then, for each grid element, we keep those with at least two RRab stars, and we take the mean derived reddening difference between the RRab stars and the RC value in that grid element. We then take the median of the comparisons for all of the grid points. 

For the LMC, using the grid of \citet{2020ApJ...889..179G}, we derive $(V-I)_{RC,0} = 0.95$ and $(V-I)_{RC,0} = 0.89$ by assuming the intrinsic (dereddened) RRab colors as derivable by Equations \ref{EQ:Catelan} and \ref{EQ:Monson}. Similarly, if we use the grid of \citet{2020arXiv200602448S}, we derive $(V-I)_{RC,0} = 0.96$ and $(V-I)_{RC,0} = 0.89$ using Equations \ref{EQ:Catelan} and \ref{EQ:Monson}. We choose the latter results from Equation \ref{EQ:Monson} as primary, as it is calibrated off of the same class of colors as it measures, the difference between intensity-weighted magnitudes. For both relationships, the median absolute deviation was 0.03 mag. The standard deviation, using Equation \ref{EQ:Monson}, when removing outlier pixels with ${\delta}E(V-I) \geq 0.15$ mag is $\sim$0.05 mag.

For the SMC, we repeat the exercise with one adjustment, we require only one RRab star per grid element, as the density of RRab stars with reliable parameters in that population appears to be lower. Using the grid of  \citet{2020ApJ...889..179G}, we derive $(V-I)_{RC,0} = 0.89$ and $(V-I)_{RC,0} = 0.82$ by assuming the intrinsic (dereddened) RRab colors as derivable by Equations \ref{EQ:Catelan} and \ref{EQ:Monson}. Similarly, if we use the grid of \citet{2020arXiv200602448S}, we derive $(V-I)_{RC,0} = 0.95$ and $(V-I)_{RC,0} = 0.87$ using Equations \ref{EQ:Catelan} and \ref{EQ:Monson}. The standard deviation, using Equation \ref{EQ:Monson}, when removing outlier pixels with ${\delta}E(V-I) \geq 0.15$ mag is $\approx$0.06 mag.

 %We derive $(V-I)_{RC,0} = 0.92$ and $(V-I)_{RC,0} = 0.84$ \textbf{by assuming the intrinsic (dereddened) RRab colors as derivable by Equations \ref{EQ:Catelan} and \ref{EQ:Monson}}. For both relationships, the median absolute deviation was 0.04 mag. The standard deviation, using Equation \ref{EQ:Monson}, when removing outlier pixels with ${\delta}E(V-I) \geq 0.15$ mag is $\sim$0.05 mag. 
 
 For both the LMC and the SMC, the scatter between $E(V-I)_{\rm{RC}}$ and $E(V-I)_{\rm{RRab}}$ is larger than that of the fit to Equation \ref{EQ:Monson}, of $\delta$=0.026 mag. This can be explained if there is some combination of a little differential reddening (${\delta}E(V-I \approx 0.05$) in each field, if the scatter to Equation \ref{EQ:Monson} is under-estimated by the small or perhaps non-representative local sample, or if the best-fit parameters to Equation \ref{EQ:Monson} are slightly off. The last possibility is explored in the next subsection. 
 
 \subsubsection{Uncertainties in the Calibration with RR Lyrae (ab-type)} \label{subsec:RRab2}
 
 The scatter in color of the 24 RRab stars from the mean relation of Equation \ref{EQ:Monson} is given by $\delta=0.026$ mag. The errors in the three coefficients are $10^{-2} (0.97,10,1.2)$, and the correlation matrix for the uncertainties is given by:
\begin{equation*}
Corr = 
\begin{pmatrix}
    1 &  0.80   & 0.66 \\
    0.80 & 1  & 0.67 \\
    0.66  & 0.67  & 1
\end{pmatrix}
\label{EQ:Correlations}
\end{equation*}

The uncertainty due to the zero point is given as 0.0097 mag. The uncertainty in the determinations of the individual periods negligibly contributes to that for individual RRab, whose periods are typically determined to 1 part in 100,000. The mean effect will be much larger, as the mean value of $\log \bigl( \rm{P}/(0.60\,\rm{days}) \bigl)$ for the \citet{2017AJ....153...96M} sample and the LMC and SMC sample differ by 0.047 and 0.059 respectively, with the latter RRab populations having, on average, longer periods. The difference in the metallicities of the LMC and SMC with respect to the sample of \citet{2017AJ....153...96M} are $0.21 \pm 0.20$ and $0.59 \pm 0.08$, with the LMC and SMC RRab stars having lower metallicities \citep{2016AcA....66..269S}.

 The relationship of Equation  \ref{EQ:Monson} necessitate a zero-point shift in the reddening maps of \citet{2020ApJ...889..179G}, to  $(V-I)_{\rm{RC,0,LMC}} \approx 0.89 \pm 0.02$ and  $(V-I)_{\rm{RC,0,SMC}} \approx 0.82 \pm 0.02$. If we repeat the exercise with the reddening maps of \citet{2020arXiv200602448S}, we derive  $(V-I)_{\rm{RC,0,LMC}} \approx 0.89 \pm 0.02$ and  $(V-I)_{\rm{RC,0,SMC}} \approx 0.87 \pm 0.02$. As the differences here are likely due to a combination of different resolution and different locations of the boundaries of each grid point, we adopt the average of these results as primary.

\subsection{Implications for the Reddening of the Tip of the Red Giant Branch in the LMC}

The Tip of the Red Giant Branch (TRGB) is a primary distance indicator that is commonly used to measure the distance to galaxies by tracing the stellar population in their halos where extinction is minimal.  Because TRGB is not a type of star but rather the maximum luminosity of many red giants it is difficult to calibrate this luminosity without reliable parallaxes for a large number of red giants near the tip \citep{2019PASA...36....1M}.  Instead, TRGB is most commonly calibrated in the LMC using its measured distance.  Unfortunately TRGB stars in the LMC suffer substantial extinction in the $I$-band and thus the calibration of TRGB in the halos depends directly on the estimate of the absolute (i.e., true, not differential) LMC extinction or reddening.  Here we consider the estimate of the LMC TRGB extinction in light of the red clump calibration from Section \ref{sec:Calibration}.

The reddening maps from \citet{2011AJ....141..158H} yield a mean value of $A_I=0.10$ (mag) for the sight lines of stars near the TRGB as determined by \citet{2017ApJ...835...28J,2017ApJ...836...74J} and \citet{2019ApJ...886...61Y} which were derived from a value of $(V-I)_{RC,0}=0.92$, consistent with the values found here.  This extinction calibration leads to $M_{I,TRGB}=-3.97 \pm 0.03$ based on the detached eclipsing binary distance to the LMC of $m-M=18.477$ \citep{2019Natur.567..200P} and an apparent $m_{I,TRGB}=14.60$.  

For an independent comparison, we measured the extinction of TRGB sight lines from LMC reddening maps derived from RR Lyrae star colors.  RR Lyrae should provide very good tracers for TRGB extinction as they are similar in age and are good calibrated crayons.  We measured the median reddening of the TRGB sight lines (TRGB samples from \citet{2017ApJ...835...28J,2017ApJ...836...74J} and \citet{2019ApJ...886...61Y}) using the RR Lyrae LMC reddening maps discussed in Section \ref{subsec:RRab} and find a median value of $A_I=0.16$ and $A_I=0.08$ mag using the \citet{2017AJ....153...96M} and \citet{2004ApJS..154..633C} relationships, respectively.  These results well match those from the \citet{2011AJ....141..158H}  red clump maps, but see further discussion in Section \ref{sec:Discussion}.

\section{Discussion} \label{sec:Discussion}

We sought to estimate the intrinsic color, $(V-I)_{RC,0}$, of the red clumps of the Magellanic Clouds. Using predictions from i) the BaSTI isochrones shifted by an offset calibrated off of three stellar populations, ii) the Cepheids period-color relations, and iii) RR Lyrae period-color relations, we have estimated that
$(V-I)_{RC,0,\rm{LMC}} = \{ \approx 0.93,0.91 \pm 0.02,0.89 \pm 0.02\}$  respectively, and using the first and third method that $ (V-I)_{RC,0,\rm{SMC}} = \{\approx 0.85,0.84 \pm 0.02 \}$  respectively. A summary of the different methods is listed in Table \ref{table:methods}. In the pursuit of that goal, we have also demonstrated the following:
\begin{enumerate}
    \item The offset between the predictions of the BaSTI v1 isochrones with Reimers' mass-loss parameter $\eta=0.20$ and the observations of red clump stars are remarkably consistent over a broad metallicity interval. The offset is given by ${\Delta}(V-I)_{RC,0},{\Delta}M_{I,RC} = (0.02 \pm 0.01,0.13 \pm 0.03)$, with the observations being redder and fainter than the theoretical predictions. That ${\Delta}(V-I)_{RC,0}$ is smaller than ${\Delta}M_{I,RC}$  is consistent with the finding of \citet{2016A&A...590A..64S}, who estimated, using the same stellar evolution code, an upper bound on the mass lost by red giant star progenitors in 47 Tuc of  ${\Delta}M \approx 0.17 M_{\odot}$, some 70\% larger than that predicted when assuming $\eta =$ 0.20. The predictions of the updated BaSTI database, BaSTI version 2.0, are both consistent with zero. In passing, this represents a sound evidence of the reliability and accuracy of the physical framework adopted for the new version of the BaSTI model library. 
    \item The different calibrations yield results that are remarkably consistent with each other. For the LMC Cepheid sample, the theoretically-calibrated period-color relation of \citet{2013MNRAS.434.2866F} is consistent  with the empirically-calibrated relation from \citet{2006ApJ...652.1133M}, from which the derived reddening estimates are consistent with those obtained using the RR Lyrae. 
\end{enumerate}

That is an impressive degree of consistency. It is also the case that reddening maps derived from observations of RR Lyrae need not perfectly agree with reddening maps derived from observations of red clump stars -- the former are a more metal-poor population, and thus likely kinematically hotter. They may have a different spatial distribution with respect to the dust. 

One calibration which is not consistent with those used in this work is that used by \citet{2020ApJ...889..179G}. They estimated  the reddening along different sightlines using the blue supergiant sample of \citet{2017AJ....154..102U} as standard crayons. A plausible explanation for this is that the blue supergiants with ages of $\sim$ 1 Myr are along sightlines with extinction greater than the mean value probed by the $\sim$3 arcminute, corresponding to approximately 44 parsecs, resolution  of the reddening maps of \citet{2020ApJ...889..179G}. Indeed, the mean value of $R_{V}=A_{V}/E(B-V)$ toward the blue supergiants is $\langle R_{V} \rangle = 4.5$. That is not consistent with the standard value of  $R_{V}=3.1$. If, for example, the reddening toward the blue supergiants is due to two components of extinction, one with $R_{V}=3.1$ and another with $R_{V}=6$, then the $R_{V}=3.1$ component, which is the only component that contributes to the red clump stars, accounts for approximately half of the extinction to the blue supergiants. Further, given that the mean error on $R_{V}$ is also approximately equal to the standard deviation in the derived $R_{V}$ values, it may be that all of the blue supergiants in the sample have this extra extinction along their line of sight. 

Similarly, the fact that the reddening estimates toward Cepheids are consistent with those toward red clump stars and RR Lyrae indicates that there is null or negligible extra extinction toward Cepheids. Whatever extra extinction there is, is likely to exist toward younger stars of $t \sim 1$ Myr, it is significantly reduced or eliminated by the time those stars are $\sim$100 Myr old, the typical age of Cepheids \citep{2005ApJ...621..966B}. For example, the study of \citet{2018ApJ...861...36A} showed that $\sim$98\% of the Cepheids in the Andromeda galaxy have migrated away from their dusty birth sites.   Because cluster disruption times are generally longer than blue supergiant ages, it is reasonable to assume additional extinction for blue supergiants comes from their birth sites. 
 
\citet{2020arXiv200602448S} derived reddening maps for the Magellanic Clouds by dereddening RC star colors in the outskirts of the Clouds (3-9 and 1-4 degrees away for the LMC and SMC, respectively) where the SFD maps \citep*{1998ApJ...500..525S} should give more valid results. They then corrected for the metallicity-dependence of the color of the red clump to derive intrinsic red clump colors of $(V-I)_{RC,0,\rm{LMC}} = 0.8915 - d \times 0.0025$, and $(V-I)_{RC,0,\rm{SMC}}=0.8816 \pm 0.0112$ for $d \leq 1.5$ degrees, and $(V-I)_{RC,0,\rm{SMC}}=0.8667 \pm 0.0031$ for $d > 1.5$

\citet{2020ApJ...891...57F} present a method to measure the reddening of TRGB stars in the LMC that does not appear to agree with the LMC reddening maps previous discussed.  They simultaneously measured the difference in distance and reddening between the LMC and another host using the colors of their respective TRGBs, e.g., in the SMC and LMC:  

\begin{equation}
\begin{split}
    \Delta \mu (\rm{SMC}-\rm{LMC})=
    (m_{x,\rm{SMC}}-R(x,I)A_{I,SMC})-\\(m_{x,\rm{LMC}}-R(x,I)A_{I,\rm{LMC}}).
\label{EQ:Freedman}
\end{split}
\end{equation}

where $m_{x,\rm{SMC}}$ is the apparent magnitude of TRGB in one of 5 bands $(V,I,J,H,K_{s})$, and $R(x,I)$ is the ratio of extinction between band $x$ and $I$.  They estimate $A_{I,\rm{SMC}}$ from the Galactic foreground reddening maps of \citet{2011ApJ...737..103S} as given in NED, then solve for two parameters, $\Delta \mu$ and $A_{I,\rm{LMC}}$.  They find a significantly higher value of $A_I=0.169 \pm 0.02$, $>$ 3 $\sigma$ higher than the value derived from the red clump star map of \citet{2011AJ....141..158H}% or the RR Lyae maps presented here and a more luminous calibration of $M_I$ for TRGB by 0.08 mag, a direct consequence of the enhanced extinction.  
However, an apparent discrepancy in measurements of the TRGB $m_{I,\rm{SMC}}$ might explain this difference.

\citet{2020ApJ...891...57F} report a value of SMC TRGB $m_{I,\rm{SMC}}=14.93 \pm 0.01$  from OGLE photometry.   \citet{2016AJ....151..167G} found the SMC TRGB from OGLE photometry is broad and brighter and varies across five SMC fields from 14.98 to 15.20 with a mean of $m_I=15.04$.  \citet{2019ApJ...886...61Y} found $m_I=15.01$ from the OGLE data, consistent with \citet{2016AJ....151..167G} but also fainter than \citet{2020ApJ...891...57F} by $\sim$ 0.1 mag.  As shown in equation \ref{EQ:Freedman}, $m_{I,\rm{SMC}}+A_{I,\rm{LMC}} \sim \Delta \mu$ so a higher value of $m_{I,\rm{SMC}}$ by $\sim$ 0.1 mag than \citet{2020ApJ...891...57F} yields a lower value of $A_{I,\rm{LMC}}$ by this difference for the same $\Delta \mu$ ($\Delta \mu$ is well constrained in the measured range by the relative detached eclipsing binary distance from \citet{2017ApJ...842..116W} with $\sigma=0.026$ mag) and could explain the discrepancy with the LMC reddening maps.   \citet{2020ApJ...891...57F} does not indicate which stars  in the SMC (or LMC) OGLE catalogues (\textit{i.e.}, the region, mag range or filtering characteristics) were employed to measure the TRGB so we cannot readily resolve or reproduce the differences in $m_{I,\rm{SMC}}$ with \citet{2016AJ....151..167G} or \citet{2019ApJ...886...61Y}.  

Another source of discrepancy is the foreground estimate of $A_{I,\rm{SMC}}$ used by \citet{2020ApJ...891...57F} via NED.  This value is quite uncertain because it can not be derived {\it directly} from the the IRAS maps \citep{1984ApJ...278L...1N}  used by \citet{2011ApJ...737..103S} because the Magellanic Clouds contaminate the foreground glow.   One may interpolate the IRAS maps around the $\sim$ 10 degree perimeter of the SMC where the maps show $\sigma \sim E(B-V) \sim$ 0.015 mag (the Magellanic Stream also provides regional contamination of the IRAS maps) resulting in an uncertainty of $A_{V,\rm{SMC}}$ of 0.05 mag which lowers the precision of the TRGB method considerably.  In addition, the large range in the SMC TRGB values may result from a large depth \citep{2012ApJ...744..128S} which also challenges the precision of the SMC to LMC comparison.  An independent comparison between the LMC and IC 1613 (old, dwarf elliptical) by \citet{2020ApJ...891...57F} avoids the issues with the SMC but the metallicities and ages of the TRGB span a larger range of one dex and systematic differences in TRGB may arise.  \citet{2020ApJ...891...57F} accounts for age and metallicity differences with a linear color term, but there is empirical evidence that this may not fully rectify the disparity; there is excessive dispersion beyond the quoted $0.01$ mag TRGB errors between hosts as the $\chi^2$ for the color comparison (equation (\ref{EQ:Freedman}) for the LMC-IC 1613 comparison) is 16 for 3 degrees of freedom (P=0.1\%).  Whether the high value of $\chi^2$ is due to higher measurement errors or intrinsic variation in color-corrected TRGB colors (see  \citet{2019ApJ...880...63M}) , the uncertainty in $A_{I,\rm{LMC}}$ would be higher by $\sim$ $\sqrt(5)$.  

Could the sight lines of TRGB in the LMC have much higher extinction than red clump or RR Lyrae tracers?  In principle, red clump and RR Lyrae stars should provide good measures of dust sight lines for near-TRGB stars as they are all highly evolved red giants of $\geq$ Gyr timescales.  In general, older populations have more time to migrate from their dusty star-forming birth sites so a hierarchy of ages would correspond to a hierachy of extinction.  As shown in Section \ref{subsec:Cepheids} the red clump star maps of \citet{2011AJ....141..158H} agree with the colors of Cepheids which are even younger (at 50-100 Myr) than the red clump stars.  TRGB stars should be a slightly older population (an older stellar population produces fewer horizontal branch and particularly fewer red clump stars per capita, and correspondingly, more red giant stars) so it is unlikely TRGB stars would have statistically more extinction than red clump stars and Cepheids.   \citet{2020ApJ...891...57F} and \citet{2019A&A...628A..51J} identify a range of recent LMC literature values of $A_I$ of 0.10 to 0.20 mag measured from B Stars, red clump Stars, RR Lyrae, star clusters, Cepheids and detached eclipsing binary stars.  The \citet{2020ApJ...891...57F} value of $A_I$=0.17 mag falls at the higher end of this range and the value of $A_I$=0.10 mag from the \citet{2011AJ....141..158H} maps at the lower end.  Because all of these stellar tracers are younger and likely along dustier sightlines than TRGB stars (with the exception of RR Lyrae), considering the range and age of the population, TRGB stars would seem likely to fall at the low end of this range.  

An attractive alternative to estimating the extinction of TRGB in the LMC is to calibrate it in the halo of the maser host NGC 4258 \citep*{2017ApJ...835...28J,2019ApJ...886L..27R}  where the result, like the application in SN Ia hosts, is insensitive to dust.

\subsection{Prospects for Further Work}
Additional and independent estimates of the reddening toward the Magellanic Clouds would be useful not just to simply further calibrate the latter, but alternatively, to explore the consistency of different approaches. Among the possibilities:
\begin{enumerate}
    \item To compare the predictions for the red clump of the BaSTI isochrones, either of version 1 or version 2, with a larger sample robust empirical anchors. We would have liked to do so, but most of the metal-rich clusters are toward the inner Milky Way and thus have non-standard extinction along their lines of sight  \citep{2008ApJ...680.1174N,2016MNRAS.456.2692N}. Separately, the horizontal branch morphologies of globular clusters are sensitive to variations in the initial abundances of helium \citep{2007A&A...474..105B,2010ApJ...708..698D,2014ApJ...785...21M,2014MNRAS.437.1609M}. The  latter are not always as well determined as they are for 47 Tuc. 
    \item To measure the foreground reddening toward the Magellanic Clouds using both a probabilistic analysis of the spectroscopic, photometric, and parallax information of foreground stars in GALAH \citep{2015MNRAS.449.2604D}, as well as the measurements of the equivalent widths of the diffuse interstellar bands in their spectra. We could not do so here as the reddening estimates are not in the most up-to-date publicly-available data release \citep{2019A&A...624A..19B}. We did attempt to do this with APOGEE data, from which we estimated the foreground  $A_{I} \approx 0.12$, but that sample was of only $\sim$30 stars. 
    \item To measure additional colors of the red clump, to break the degeneracy between reddening variations and intrinsic color variations. We attempted to do this with the $(g-i)$ measurements of \citet{2018ApJ...866...90C}, but this did not yield a robust constraint as $E(g-i)/E(V-I)$ has approximately the same value as the $\delta(g-i)/\delta (V-I)$ that results from variations in stellar effective temperature. In contrast, with infra-red data, $E(I-K_{s})/E(V-I) \approx 1.04$, and $\delta(I-K_{s})/\delta (V-I) \approx 0.47$.  
    \item To use open and globular clusters in the Magellanic Clouds as standard crayons. At this time, that literature is very heterogeneous in its assumptions (choice of isochrones, rotation rates, etc) and thus the resultant reddening scale ends up being subjective. Further, many of the clusters have photometric metallicity estimates, for which the derivation is degenerate with that of the reddening. 
    \item To obtain spectroscopic [Fe/H] and hopefully [$\alpha$/Fe] measurements for a larger number of solar neighbourhood RR Lyrae. The correlations in the errors to the terms of Equation \ref{EQ:Monson} are all large and positive, $+0.66$, $+0.67$, and $+0.80$. These could be reduced, and better understand, with a larger sample of RR Lyrae stars with robustly-determined metallicities. Ideally, some LMC and SMC RR Lyrae would have their spectroscopic abundances measured in a consistent manner to those of the solar neighbourhood. An investigation of the difference between static magnitudes, time-weighted mean magnitudes, and intensity-weighted mean magnitudes of RR Lyrae for every bandpass would also be beneficial. 
\end{enumerate}

%Separately of the colors, our estimates of $M_{I,RC,\rm{LMC}}=-0.33$ and $M_{I,RC,\rm{SMC}}=-0.38$ are further validation that an issue that once challenged the literature is indeed resolved. \citet{1998AcA....48..383} measured the deredenned brightness of the red clumps of the Magellanic Clouds, and estimated that $(m-M)_{\rm{LMC}}=18.18 \pm 0.06$ mag and $(m-M)_{\rm{SMC}}=18.65 \pm 0.08$ 

\section{Conclusion} \label{sec:Conclusion}

In this investigation we have investigated three different methods to estimate the intrinsic color, $(V-I)_{RC,0}$ of the red clumps of the inner few degrees of the Magellanic Clouds. 

The first method was derived from the predictions of the BaSTI version 1 isochrones, for which we applied empirically-calibrated offsets calibrated by comparing the isochrone predictions to the observations of the globular cluster 47 Tuc, the solar neighbourhood, and the open cluster NGC 6791. Together, these three populations constitute a robust calibration set, as they span a metallicity range of ${\Delta}$[Fe/H]$\approx 1.1$ dex. The BaSTI isochrones predict red clumps that are slightly red and bright relative to the observations, by ${\Delta}(V-I)_{RC,0},{\Delta}M_{I,RC} \approx (0.02,0.13)$, with these offsets being almost equal to the offsets between the BaSTI version 1 and version 2 isochrones.  We applied this offset to calculate our first estimate of the red clumps of the Magellanic Clouds, and obtained $(V-I)_{RC,0,\rm{LMC}}=0.93$ and $(V-I)_{RC,0,\rm{SMC}}=0.85$, and also $M_{I,RC,\rm{LMC}} \approx -0.26$ and $M_{I,RC,\rm{SMC}} \approx -0.37$.

The second method derived estimates by comparing the reddening toward Cepheids, for which we assumed that the average intrinsic colors can be reliably determined from period-color relations, with the reddening values estimated from the nearest red clump centroid measurement. The estimate was nearly the same regardless of whether or not we used the empirically-calibrated period-color relation of \citet{2006ApJ...652.1133M} or the theoretically-derived period-color relation of \citet{2013MNRAS.434.2866F}. The resulting estimate was $(V-I)_{RC,0,\rm{LMC}}=0.91 \pm 0.02$.

The third method derived estimates by using the OGLE RR Lyrae catalogue of \citet{2016AcA....66..131S}, with metallicity estimates from \citet{2016AcA....66..269S} as a set of standard crayons with which to estimate the reddening along their lines of sight. The $(V-I)_{\rm{RRab}}$ colors predicted by the empirically calibrated period-color relation from the data  of \citet{2017AJ....153...96M} yields the the estimates $(V-I)_{RC,0,\rm{LMC}}=0.89 \pm 0.02$ and $(V-I)_{RC,0,\rm{SMC}}=0.84 \pm 0.02$.

%With these three methods, we estimated the intrinsic color of the red clump of the Large Magellanic Cloud to be $(V-I)_{RC,0,\rm{LMC}} = \{0.90,0.93,0.93\}$ respectively, and with the first and third method we estimated the intrinsic color of the red clump of the Small Magellanic Cloud to be $ (V-I)_{RC,0,\rm{SMC}} = \{0.85,0.88 \}$ respectively. These color zeropoints can be used to improve the calibration of existing Magellanic Cloud reddening maps and distance scale measurements.

\section*{Acknowledgments}

We thank the referee for helpful feedback on the manuscript. 

We thank Aaron Dotter, Adriano Pietrinferni, Dan Scolnic, Andrea Kunder, Marcio Catelan, Dan Weisz, Marek Gorski, and Dorota Skowron for helpful discussion. 
	
DMN acknowledges support from NASA under award Number 80NSSC19K0589, and support from the  Allan C. And Dorothy H.Davis Fellowship. SC acknowledges support
from Premiale INAF MITiC, from Istituto Nazionale di Fisica
Nucleare (INFN) (Iniziativa specifica TAsP), and grant AYA2013-
42781P from the Ministry of Economy and Competitiveness of
Spain.
LC acknowledges support from the Australian Research Council Future Fellowship FT160100402.

This work has made use of data from the European Space Agency (ESA) mission
{\it Gaia} (\url{https://www.cosmos.esa.int/gaia}), processed by the {\it Gaia}
Data Processing and Analysis Consortium (DPAC,
\url{https://www.cosmos.esa.int/web/gaia/dpac/consortium}). Funding for the DPAC
has been provided by national institutions, in particular the institutions
participating in the {\it Gaia} Multilateral Agreement.

This publication makes use of data products from the Two Micron All Sky Survey, which is a joint project of the University of Massachusetts and the Infrared Processing and Analysis Center/California Institute of Technology, funded by the National Aeronautics and Space Administration and the National Science Foundation.

Funding for the Sloan Digital Sky Survey IV has been provided by the Alfred P. Sloan Foundation, the U.S. Department of Energy Office of Science, and the Participating Institutions. SDSS-IV acknowledges
support and resources from the Center for High-Performance Computing at
the University of Utah. The SDSS web site is www.sdss.org.

SDSS-IV is managed by the Astrophysical Research Consortium for the 
Participating Institutions of the SDSS Collaboration including the 
Brazilian Participation Group, the Carnegie Institution for Science, 
Carnegie Mellon University, the Chilean Participation Group, the French Participation Group, Harvard-Smithsonian Center for Astrophysics, 
Instituto de Astrof\'isica de Canarias, The Johns Hopkins University, Kavli Institute for the Physics and Mathematics of the Universe (IPMU) / 
University of Tokyo, the Korean Participation Group, Lawrence Berkeley National Laboratory, 
Leibniz Institut f\"ur Astrophysik Potsdam (AIP),  
Max-Planck-Institut f\"ur Astronomie (MPIA Heidelberg), 
Max-Planck-Institut f\"ur Astrophysik (MPA Garching), 
Max-Planck-Institut f\"ur Extraterrestrische Physik (MPE), 
National Astronomical Observatories of China, New Mexico State University, 
New York University, University of Notre Dame, 
Observat\'ario Nacional / MCTI, The Ohio State University, 
Pennsylvania State University, Shanghai Astronomical Observatory, 
United Kingdom Participation Group,
Universidad Nacional Aut\'onoma de M\'exico, University of Arizona, 
University of Colorado Boulder, University of Oxford, University of Portsmouth, 
University of Utah, University of Virginia, University of Washington, University of Wisconsin, 
Vanderbilt University, and Yale University.

\bibliography{NatafRiessLMCSMC}

\end{document}